\documentclass[aps,twocolumn,pra,superscriptaddress,amsmath,showpacs,tightenlines,nofootinbib]{revtex4}
\usepackage{epsfig,graphicx,times}
\usepackage{amstext}
\usepackage{amsmath}            
\usepackage{amssymb}            
\usepackage{graphicx}           
\usepackage{latexsym}
\usepackage{bm}
\usepackage{color}
\begin{document}
\title{Mode coupling and photon antibunching in a bimodal cavity containing a dipole-quantum-emitter}
\author{Yu-Long Liu}
\affiliation{Institute of Microelectronics, Tsinghua University, Beijing 100084, China}
\affiliation{Tsinghua National Laboratory for Information Science and Technology (TNList),
Beijing 100084, China}
\affiliation{CEMS, RIKEN, Saitama 351-0198, Japan}
\author{Guan-Zhong Wang}
\affiliation{Institute of Microelectronics, Tsinghua University, Beijing 100084, China}
\author{Yu-xi Liu}
\email{yuxiliu@mail.tsinghua.edu.cn}
\affiliation{Institute of Microelectronics, Tsinghua University, Beijing 100084, China}
\affiliation{Tsinghua National Laboratory for Information Science and Technology (TNList),
Beijing 100084, China}
\affiliation{CEMS, RIKEN, Saitama 351-0198, Japan}
\author{Franco Nori}
\affiliation{CEMS, RIKEN, Saitama 351-0198, Japan}
\affiliation{Department of Physics, The University of Michigan, Ann Arbor, Michigan
48109-1040, USA}
\date{\today }

\begin{abstract}
We study the effect of mode-coupling on a single-photon device in which a
dipole-quantum-emitter (DQE) is embedded in a bimodal whispering-gallery-mode
cavity (WGMC). A scatterer is used to induce mode coupling between
counter-clockwise and clockwise propagating light fields, which interact with
the DQE. In contrast to models for the interaction between a DQE
and a (one-mode or two-mode) cavity field, we find that strong photon
antibunching can occur even for a weak DQE-field coupling and large
dephasing of the DQE, when mode coupling is introduced. We also find that mode
coupling can make the device robust against either the frequency mismatch
between cavity modes and the DQE or the coupling strength mismatch between
the DQE and each mode in the two-mode cavity. Moreover, we find that these
mismatches can be used to generate better antibunching in the weak
DQE-field coupling regime. Our study shows that mode coupling in a
bimodal cavity is very important for the realization of a good single-photon device.

\end{abstract}

\maketitle

\pagenumbering{arabic}

\section{Introduction}

Single-photon devices play a crucial role in quantum information science,
including quantum communication and quantum computing technology~\cite{1,2}.
These devices are useful for quantum key distribution~\cite{3,4}, generation
of entangled states~\cite{5}, quantum metrology~\cite{6,7,8,9}, single-photon
quantum memory~\cite{10,11,12,13}, linear optical quantum
computing~\cite{14,15}, and quantum simulations~\cite{buluta1,buluta2}. Thus, a highly-efficient method for generating
single-photons with low error is urgently required for quantum information
processing. Similar to Coulomb blockade for electrons in mesoscopic electronic
devices~\cite{16,17,18}, photon blockade, where the subsequent photons are
prevented from resonantly entering a cavity, is a promising way to produce
single-photon sources. Photon blockade requires that the strength of the
photon-photon Kerr nonlinear interaction~\cite{19,20} is much larger than the
decay rate of the cavity field.

In principle, the photon blockade can be measured by the second-order
correlation functions of photons~\cite{correlation measurements,21}. Photon
blockade in the optical frequency domain has been observed in: a trapped atom
coupled to a cavity field~\cite{22}, a two-level system coupled to a bimodal
microtoroidal resonator~\cite{23}, and a quantum dot coupled to single-mode
field of a photonic crystal resonator~\cite{24}. Recently, photon blockade in
the microwave frequency domain was also observed~\cite{25,26} by coupling a
superconducting artificial atom~\cite{review1,review2,review3} to a
transmission line resonator. A crucial prerequisite for photon blockade in
these experiments~\cite{22,23,24,25,26} is to reach the strong field-atom
coupling regime~\cite{27}, i.e., the coupling strength between the cavity
field and the dipole-quantum-emitter (DQE) exceeding the decay rates of both
the cavity field and DQE. The strong DQE-field coupling can induce a strong nonlinear
photon-photon interaction and could be used to produce photon blockade
and single-photon sources~\cite{L,X,Liao1,Liao2,Lv}. However, when a cavity
field works at the single-photon level, it is not easy to achieve a
cavity-atom-induced strong photon-photon interaction.

Recently,  photonic \textquotedblleft molecule\textquotedblright\  systems,
consisting of two coupled cavities with Kerr nonlinearities, were proposed to
generate antibunching photon and photon blockade~\cite{28}. In contrast to
previous studies~\cite{22,23,24,25,26}, it was found that the photon blockade
can occur even in a weak field-atom coupling regime, when an additional cavity
is coupled~\cite{29}. This new mechanism corresponds to a destructive quantum
interference effect in nonlinear photonic molecule systems~\cite{29}.
Motivated by several studies~\cite{28,29}, various systems were proposed to
achieve photon blockade and antibunching photons, such as coupled
optomechanical systems~\cite{30,31,Ludwig-PRL2012}, dipolariton systems formed
by cavity fields and excitons~\cite{32}, coherent-feedback controlled
optomechanical systems~\cite{33}, and coupled microcavities with second- or
third-order nonlinearities~\cite{34,35}. However, the role of the coupled
cavity was not explored in detail. Thus it is interesting to further explore
this issue as a function of the coupling strength between the two cavities.

Motivated by an experiment~\cite{23} on a photon turnstile, in which a DQE
is embeded in a bimodal whispering-gallery-mode
cavity (WGMC)~\cite{23}, we here
mainly study the mode coupling effect on the nonclassical properties of
photons. For two strongly-coupled
resonators~\cite{28,29,30,31,Ludwig-PRL2012,32,33,34,35}, two modes of two
resonators are coupled to different nonlinear quantum systems or only one of
the cavity modes is coupled to a nonlinear quantum system. However, here: (i)
a single WGMC is used to support two degenerate counter-propagating cavity
modes; (ii) the mode coupling or normal-mode splitting between these two
counter-propagating modes is realized by a scattering
object~\cite{36,37,38,39,40,41,42,43,44}; (iii) two cavity modes are
coupled to the same DQE. Compared to the studies where a two-level system is
coupled to two modes of a cavity~\cite{45,46}, here the mode coupling is
introduced by a scattering object.

The remaining part of the paper is organized as follows. In Sec.~II, the
theoretical model is introduced. In Sec.~III, the measurements for the
mode-coupling are discussed. In Sec.~IV, the photon blockade is analytically
and numerically studied in different parameter regimes via the second-order
correlation function. In particular, we analyze the photon blockade in the
weak DQE-field coupling regime. We also compare our results with those in
coupled-cavity systems~\cite{28,29,30,31,Ludwig-PRL2012,32,33,34,35}. In
particular, we analyze the physical mechanism of the photon blockade in the
system studied here. In Sec.~V, we study the robustness of the system on the
dephasing of the DQE. In Sec.~VI, we study the robustness of the system on
either the frequency mismatches between the cavity mode and the DQE or the
coupling strength mismatch between the DQE and each cavity mode.
Conclusions and perspective discussions are finally presented in
Sec.~VII.

\begin{figure}[ptb]
\includegraphics[bb=125 0 650 580, width=8cm, clip]{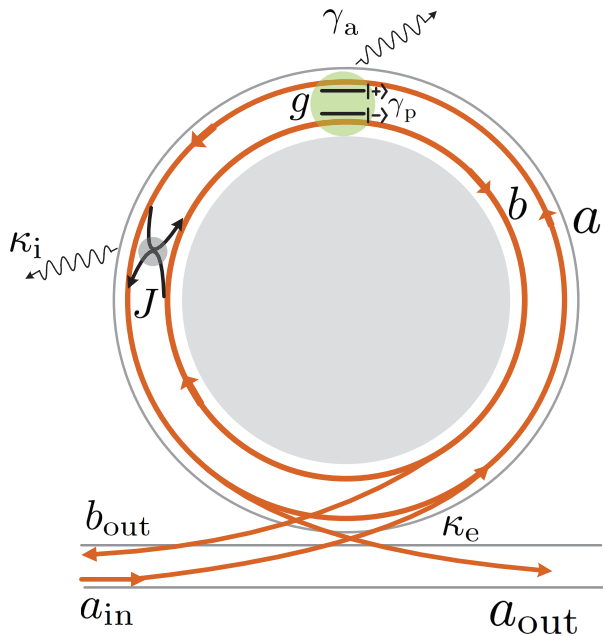} \caption{(Color
online) Schematic of a WGMC-DQE system which is driven via an optical fiber
taper waveguide with a cavity-waveguide coupling rate $\kappa_{\mathrm{e}}$.
The nanoparticle in the WGMC, causes an effective
coupling at a rate $J$, between the clockwise and counterclockwise propagating
modes. Such two WGM modes have a DQE-field coupling rate $g$ and intrinisc
cavity decay rate $\kappa_{\mathrm{i}}$. The DQE has a spontaneous emission
rate $\gamma_{\mathrm{a}}$ and a phase dephasing rate $\gamma_{\mathrm{p}}.$}%
\label{Figure1}%
\end{figure}

\section{Hamiltonian}

As schematically shown in Fig.~\ref{Figure1}, we study a system that consists
of a single whispering-gallery-mode cavity (WGMC) and a dipole-quantum-emitter (DQE). As studied in Ref.~\cite{23}, the WGMC supports a
counterclockwise (CCW) and a clockwise (CW) propagating modes which have the
same frequency. These two modes are coupled to each other through a nanoparticle (called as a scatterer) The DQE is modeled as a two-level system, which can be a quantum-dot, an atom, or other systems. Here we do not
focus on any particular system, but consider a generic DQE. The Hamiltonian of
the whole system is given by~\cite{23,43,44}
\begin{align}
H  &  =\hbar\omega\left(  a^{\dag}a+ b^{\dag}b\right)  +\hbar\omega
_{\mathrm{a}}\sigma_{z}+\hbar J(a^{\dag}b+b^{\dag}a)\nonumber\\
&  +\hbar\left[  g_{a}a^{\dag}\sigma_{-}+g_{b}b^{\dag}\sigma_{-}+\varepsilon
a^{\dag}e^{-i\omega_{\mathrm{d}}t}+\text{H.c.}\right]  , \label{H1}%
\end{align}
where $a$ and $b$ ($a^{\dag}$ and $b^{\dagger}$) are the annihilation
(creation) operators of CCW and CW propagating modes with frequency $\omega$,
respectively. $\sigma_{-}$ and its conjugate operator $\sigma_{+}$ are
the ladder operators describing a DQE with frequency $\omega_{a}$, and $J$ is
the mode-coupling strength between the CCW and CW modes. We assume that the DQE and
two propagating modes have coupling strengths $g_{a}$ and $g_{b}$. The driving field is assumed to be coupled to the CCW mode with the coupling strength $\varepsilon$.

The mode coupling in the WGMC is usually introduced when a scatterer is
present. The scatterer can be considered as a nanoparticle, which is used to
enhance~\cite{38} and control~\cite{39} the effective coupling strength $J$
between the CCW and CW modes. Let us consider a simple example where the
nanoparticle is assumed to be a nanosphere with radius $R\ll\lambda$, where
$\lambda$ is the wavelength of the light field. Then the response of a
particle to an electromagnetic field can be calculated by the
Clausius-Mossotti relation and the effective coupling strength $J$
becomes~\cite{42}
\begin{equation}
J=-\frac{1}{2V}\alpha f^{2}\omega,\label{E1}%
\end{equation}
in the electrostatic limit, with%
\begin{equation}
\alpha=4\pi R^{3}\left(  \frac{n_{\mathrm{p}}^{2}-1}{n_{\mathrm{p}}^{2}%
+2}\right)  .\label{E2}%
\end{equation}
Here $f$ is the mode function of the WGMC, $V$ is the mode volume of the WGMC,
$n_{p}$ is the refractive index of the particle, and the surrounding medium is
assumed to be air.

The scatterer-induced mode splitting has been observed experimentally, and
proposed for a highly sensitive and robust platform for detecting nanoscale
objects~\cite{36,37,38,39,40,41,42}. As reported in~\cite{36}, a nanoparticle
of radius 40 nm can introduce an apparent mode splitting with a mode-coupling
strength $J>\kappa$, where $\kappa$ is the decay rate of the cavity field,
producing a strong-mode coupling. From Eqs.~(\ref{E1}) and (\ref{E2}), it is
clear that the mode-coupling strength $J$ can be enhanced by increasing the
radius of the nanoparticle. The effective coupling strength can also be
enhanced by increasing the number of nanoparticles~\cite{42}, improving the
quality factor of the microresonator~\cite{38}, introducing gain medium into
the cavity~\cite{41}, or using Raman-gain-induced loss
compensation~\cite{Raman-gain}. Here, we will not pay more attention to how
the nanoparticle will affect the system. Instead, we just assume that the
mode-coupling strength $J$ is a positive and real number, which can be controlled
and set to be much larger than the decay rate $\kappa$ of the cavity field. We
will mainly focus on the study of how the mode-coupling strength $J$ affects
the single-photon behavior in the system.

In the rotating reference frame at the driving-field frequency $\omega
_{\mathrm{d}}$, i.e., with a unitary transformation $U=\exp\left[
-i\omega_{\mathrm{d}}(a^{\dag}a+b^{\dag}b+\sigma_{z})t\right]  $, the
Hamiltonian in Eq.~(\ref{H1}) is given as
\begin{align}
H_{\mathrm{eff}} &  =\hbar\Delta \left(a^{\dag}a+ b^{\dag}b\right)+\hbar
\Delta_{\mathrm{a}}\sigma_{z}+\hbar J(a^{\dag}b+b^{\dag}a)\nonumber\\
&  +\hbar\left[  g(a^{\dag}+b^{\dagger})\sigma_{-}+\varepsilon a^{\dag
}+\text{H.c.}\right]  ,\label{H2}%
\end{align}
with the detunings $\Delta=\omega-\omega_{\mathrm{d}}$ and $\Delta
_{\mathrm{a}}=\omega_{\mathrm{a}}-\omega_{\mathrm{d}}$, from the driving field
$\omega_{\mathrm{d}}$ to the cavity modes $\omega$ and to the DQE
$\omega_{\mathrm{a}}$. Below, we first discuss the case for $g_{\mathrm{a}%
}\equiv g_{\mathrm{b}}\equiv g$ and $\Delta_{\mathrm{a}}=\Delta$ (i.e.,
$\omega=\omega_{\mathrm{a}}$) in Secs.~III, IV, and V. The mismatch of the coupling strengths
$g_{\mathrm{a}}$ and $g_{\mathrm{b}}$ and the mismatch of the frequencies
$\omega$ and $\omega_{\mathrm{a}}$ will be discussed in
Sec.~VI.

\begin{figure}[ptb]
\includegraphics[bb=0 197 584 603, width=8cm, clip]{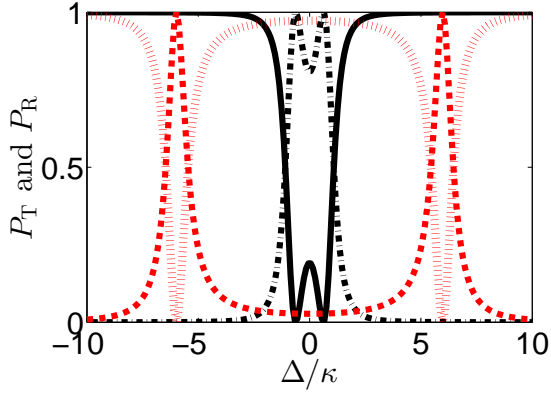} \caption{(Color
online) Normalized transmission spectra $P_{\mathrm{T}}$ and reflection
spectra $P_{\mathrm{R}}$ of the coupled WGMC-mode system. The horizontal axis
shows the normalized $\Delta=\omega-\omega_{\mathrm{d}}$, which is the
detuning from the driving frequency $\omega_{\mathrm{d}}$. Also, $\kappa$\ is
the decay rate of the cavity field. Two different parameter regimes are shown:
(i) when the optical mode-coupling rate $J$ is much larger than the
cavity-field linewidth $\kappa$, corresponding to the red-dotted curve for
$P_{\mathrm{T}}$ and red-dashed curve for $P_{\mathrm{R}}$, with $J=6\,\kappa$
as an example; \ and also (ii) when $J$ is comparable to $\kappa$,
corresponding to the black-solid curve for $P_{\mathrm{T}}$ and
black-dash-dotted curve for $P_{\mathrm{R}}$, with $J=0.8\,\kappa$ as an
example.}%
\label{Figure2}%
\end{figure}

\section{Measurement of the mode-coupling strength}

Let us first study how the mode-coupling strength $J$ could be measured. The
dynamics of the whole system is described by the quantum Langevin equations%
\begin{align}
\frac{da}{dt} &  =-\left(  i\Delta+\frac{\kappa}{2}\right)  a-iJb-ig\sigma
_{-}-i\varepsilon-\sqrt{\kappa}\,a_{\mathrm{in}},\label{eq:5}\\
\frac{db}{dt} &  =-\left(  i\Delta+\frac{\kappa}{2}\right)  b-iJa-ig\sigma
_{-}-\sqrt{\kappa}\,b_{\mathrm{in}},\label{eq:6}\\
\frac{d\sigma_{-}}{dt} &  =-\left(  i\Delta+\frac{\gamma}{2}\right)
\sigma_{-}+iga\sigma_{\mathrm{z}}+igb\sigma_{\mathrm{z}}-\sqrt{\gamma}%
\,\sigma_{\mathrm{in}}.\label{eq;7}%
\end{align}
Here, the decay rates of the two cavity modes are denoted by $\kappa$,
which consists of the intrinsic loss $\kappa_{\mathrm{i}}$ and the external
loss $\kappa_{\mathrm{e}}$, i.e., $\kappa=\kappa_{\mathrm{i}}+\kappa
_{\mathrm{e}}$. For the DQE, the decay rate $\gamma$ consists of the
spontaneous emission rate $\gamma_{a}$ and phase dephasing rate $\gamma
_{\mathrm{p}}$, i.e., $\gamma=\gamma_{\mathrm{a}}+2\gamma_{\mathrm{p}}$. Here,
$a_{\mathrm{in}}$, $b_{\mathrm{in}}$ and $\sigma_{\mathrm{in}}$ are the noise
operators associated with the CCW mode, CW mode, and the DQE with zero
mean-value, i.e., $\left\langle a_{\mathrm{in}}\right\rangle =\left\langle
b_{\mathrm{in}}\right\rangle =\left\langle \sigma_{\mathrm{in}}\right\rangle
=0$.

The mode coupling can be clearly observed in the transmitted ($P_{\mathrm{T}}%
$) and reflected ($P_{\mathrm{R}}$) cavity field power. Let us now calculate
these through the outputs of the modes $a$ and $b$, respectively. Using the
mean field approximation and from Eqs.~(\ref{eq:5}) and (\ref{eq:6}), the
equations of motion for the mean value of each operator can be given as%
\begin{align}
\frac{d}{dt}\left\langle a\right\rangle  &  =-\left(  i\Delta+\frac{\kappa}%
{2}\right)  \left\langle a\right\rangle -iJ\left\langle b\right\rangle
-ig\left\langle \sigma_{-}\right\rangle -i\varepsilon,\\
\frac{d}{dt}\left\langle b\right\rangle  &  =-\left(  i\Delta+\frac{\kappa}%
{2}\right)  \left\langle b\right\rangle -iJ\left\langle a\right\rangle
-ig\left\langle \sigma_{-}\right\rangle ,\\
\frac{d}{dt}\left\langle \sigma_{-}\right\rangle  &  =-\left(  i\Delta
+\frac{\gamma}{2}\right)  \left\langle \sigma\right\rangle +ig(\left\langle
a\right\rangle +\left\langle b\right\rangle )\left\langle \sigma_{\mathrm{z}%
}\right\rangle .
\end{align}
Using the input-output theory~\cite{Input,Output}, the output of each mode can
be obtained as%
\begin{align}
\left\langle a_{\mathrm{out}}\right\rangle  &  =\frac{i\varepsilon}%
{\sqrt{\kappa}}+\sqrt{\kappa}\,\left\langle a\right\rangle ,\label{a1}\\
\left\langle b_{\mathrm{out}}\right\rangle  &  =\sqrt{\kappa}\,\left\langle
b\right\rangle .\label{a2}%
\end{align}
Here, we are only interested in the mode coupling induced by the scatterer, thus we can
set $g=0$. Then the dynamical equations of the cavity modes are simplified
to%
\begin{align}
\frac{d}{dt}\left\langle a\right\rangle  &  =-\left(  i\Delta+\frac{\kappa}%
{2}\right)  \left\langle a\right\rangle -iJ\left\langle b\right\rangle
-i\varepsilon,\label{steady}\\
\frac{d}{dt}\left\langle b\right\rangle  &  =-\left(  i\Delta+\frac{\kappa}%
{2}\right)  \left\langle b\right\rangle -iJ\left\langle a\right\rangle.\label{steady2}%
\end{align}
By solving Eqs.~(\ref{steady}) and (\ref{steady2}) in the steady-state with
$\langle\dot{a}\rangle=\langle\dot{b}\rangle=0$, we have%
\begin{align}
\left\langle a\right\rangle  &  =\frac{\Delta-i\frac{\kappa}{2}}{\left(
i\Delta+\frac{\kappa}{2}\right)  ^{2}+J^{2}}\varepsilon,\\
\left\langle b\right\rangle  &  =\frac{-J}{\left(  i\Delta+\frac{\kappa}%
{2}\right)  ^{2}+J^{2}}\varepsilon.
\end{align}
Combining Eqs.~(\ref{a1}) and (\ref{a2}) with the steady-state solutions, the
normalized transmission power $P_{\mathrm{T}}$ and reflection power
$P_{\mathrm{R}}$ from the WGMC are given by
\begin{align}
P_{\mathrm{T}} &  =\frac{1}{\varepsilon^{2}}\left\vert \left\langle
a_{\mathrm{out}}\right\rangle \right\vert ^{2}=\left\vert \frac{i}%
{\sqrt{\kappa}}+\frac{\sqrt{\kappa}\,\left(  \Delta-i\frac{\kappa}{2}\right)
}{\left(  i\Delta+\frac{\kappa}{2}\right)  ^{2}+J^{2}}\right\vert ^{2},\\
P_{\mathrm{R}} &  =\frac{1}{\varepsilon^{2}}\left\vert \left\langle
b_{\mathrm{out}}\right\rangle \right\vert ^{2}=\left\vert \frac{\sqrt{\kappa
}\,J}{\left(  i\Delta+\frac{\kappa}{2}\right)  ^{2}+J^{2}}\right\vert ^{2}.
\end{align}

In Fig.~\ref{Figure2}, the transmission $P_{\mathrm{T}}$ and reflection
$P_{\mathrm{R}}$ powers are plotted as a function of the detuning parameter
$\Delta=\omega-\omega_{\mathrm{d}}$. Two different parameter regimes are
discussed: (i) $J>\kappa$ and (ii) $J\sim\kappa$. For $J>\kappa$, it is
obvious that a pair of resonances appear and are located at $\Delta\pm J$.
Actually, these resonances correspond to two different standing light waves,
which are superpositions (i.e. $(a+b)/\sqrt{2}$ and $(a-b)/\sqrt{2}$) of the
two counter-propagating cavity modes. When the coupling strength $J$ is
much larger than $\kappa$, a well-resolved mode-splitting can be clearly
observed in the transmission or reflection spectra. However, the distance
between two resonances becomes close when $J\sim\kappa$, and is no longer
distinguishable when $J\ll\kappa$.

\section{Photon blockade in the weak DQE-field coupling regime}

In this section we will study the photon blockade. We mainly study how the quantum behavior
of the mode fields in the WGMC varies with the change of the mode-coupling strength
$J$ between the two modes. Compared with previous studies on the strong coupling
between a two-level atom and a single-mode cavity (e.g., in Refs.~\cite{22,23,24,25,26}) or a bimodal cavity without the mode coupling~\cite{45,46}, we will here focus on the weak coupling between a DQE and the cavity modes with mode coupling.

\subsection{Master equation and second-order correlation functions}
By taking the dissipation into account, we can write the master
equation~\cite{47,48,49} of the reduced density matrix $\rho$ for whole system
as%
\begin{align}
\frac{d\rho}{dt} &  =\frac{1}{i\hbar}[H_{\mathrm{eff}},\rho]+\frac{\kappa}%
{2}(2a\rho a^{\dag}-a^{\dag}a\rho-\rho a^{\dag}a)\nonumber\\
&  +\frac{\kappa}{2}(2b\rho b^{\dag}-b^{\dag}b\rho-\rho b^{\dag}b)\nonumber\\
&  +\frac{\gamma_{\mathrm{a}}}{2}(2\sigma\rho\sigma^{\dag}-\sigma^{\dag}%
\sigma\rho-\rho\sigma^{\dag}\sigma)\nonumber\\
&  +\frac{\gamma_{\mathrm{p}}}{2}(2\sigma_{z}\rho\sigma_{z}-\rho
),\label{Master1}%
\end{align}
in the Markov approximation. Here, $H_{\mathrm{eff}}$ is given in
Eq.~(\ref{H2}). We assume that the cavity fields and the DQE are in a
zero-temperature environment, to simplify the calculation. In the number states
basis $\left\vert n_{\mathrm{a}},n_{\mathrm{b}},i\right\rangle $ and
$\left\vert n_{\mathrm{a^{\prime}}},n_{\mathrm{b^{\prime}}},i^{\prime
}\right\rangle $, the formal solution of $\rho$ in Eq. (\ref{Master1}) can be
given as \cite{28}:%
\begin{equation}
\rho(t)=\sum_{n_{a},n_{b}}\sum_{n_{a^{\prime}},n_{b^{\prime}}}\sum
_{i,i^{\prime}}\rho_{n_{a},n_{b},i;n_{a^{\prime}},n_{b^{\prime}},i^{\prime}%
}\left\vert n_{a},n_{b},i\right\rangle \left\langle n_{a^{\prime}%
},n_{b^{\prime}}i^{\prime}\right\vert .\label{Master4}%
\end{equation}
where $n_{a}\,(n_{a^{\prime}})$ represents the photon number of the CCW mode,
$n_{b}\,(n_{b^{\prime}})$ represents the photon number of the CW mode and
$i\,(i^{\prime})=+,-$ represents the excited and ground state of the DQE,
respectively.

\begin{figure}[ptb]
\includegraphics[bb=11 175 590 660,  width=0.5\textwidth, clip]{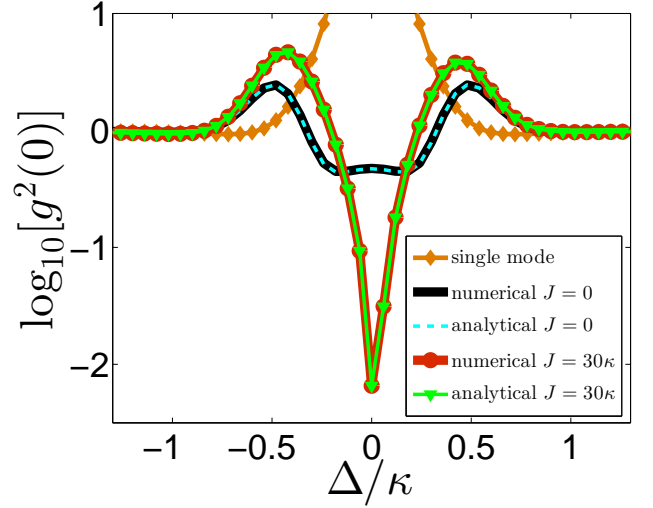}
\caption{(Color online) The second-order correlation functions $g_{\mathrm{a}%
}^{\mathrm{(2)}}(0)$ of a CCW propagating intercavity field versus the detuning
parameter $\Delta/\kappa=(\omega-\omega_{\mathrm{d}})/\kappa$, for a number of
different parameters. These correlations are calculated numerically using the
master equation and also analytically using the Schr\"{o}edinger equation in the
steady-state. The $g_{\mathrm{a}}^{\mathrm{(2)}}(0)$ is shown as an orange-rhombus solid curve for the interaction between a DQE and a single-mode cavity field; The $g_{\mathrm{a}}^{\mathrm{(2)}}(0)$ is shown as a black solid curve (numerically calculated) and a
cyan-dotted line (analytically calculated) for the interaction between a DQE and two-mode cavity fields with $J=0$. The $g_{\mathrm{a}}^{\mathrm{(2)}%
}(0)$ is shown as a red-circular solid curve (numerically calculated) and a green-triangle solid
curve (analytically calculated) for the interaction between a DQE and two-mode cavity fields with $J=30\kappa$. The system parameters for this simulation
are: $\kappa=40\gamma_{\mathrm{a}}$, $g=20\gamma_{\mathrm{a}}$, $\varepsilon
=\gamma_{\mathrm{a}}$, and $\gamma_{\mathrm{a}}=1$ MHz.}%
\label{Figure3}%
\end{figure}

We can obtain the steady-state solution $\rho_{ss}$ of the reduced density
operator $\rho$ by setting $d\rho/dt=0$, and then the statistical properties
of the driven CCW propagating cavity mode in the WGMC can be studied via the
normalized equal-time second-order correlation function%
\begin{equation}
g_{\mathrm{CCW}}^{\mathrm{(2)}}(0)=\frac{\left\langle a^{\dag}a^{\dag
}aa\right\rangle }{\left\langle a^{\dag}a\right\rangle ^{\mathrm{2}}}%
=\frac{\mathrm{Tr}(\rho_{\mathrm{ss}}a^{\dag}a^{\dag}aa)}{[\mathrm{Tr}%
(\rho_{\mathrm{ss}}a^{\dag}a)]^{\mathrm{2}}}.\label{Master}%
\end{equation}
In this section, we first study the antibunching effect of the CCW mode by
setting $\gamma_{p}=0$, in order to consider the same environmental effect as
in previous studies~\cite{23}. The effect of dephasing will be studied in Sec.~V. We
point out that the statistical properties of the CW mode can also be studied
in a similar way as those for the CCW mode.

\subsection{Numerical calculations}

We first study how the second-order correlation functions $g_{\mathrm{CCW}%
}^{\mathrm{(2)}}(0)$ vary with the detuning $\Delta=\omega-\omega_{\mathrm{d}%
}$ between the driving field and the cavity modes, and also show how the
statistical properties of the cavity field are changed by the mode-coupling
strength $J$. In Fig.~\ref{Figure3}, $g_{\mathrm{CCW}}^{\mathrm{(2)}}(0)$
versus the detuning $\Delta$ is plotted in the DQE-field weak-coupling
regime for a DQE interacting with either: (i) a single-mode cavity field, or (ii)
bimodal cavity fields without mode coupling, or (iii) bimodal cavity
fields with mode coupling. Figure~\ref{Figure3} clearly shows that there is
only photon bunching when only a single-mode cavity field is coupled to the
DQE. This is due to the weak DQE-field coupling, i.e., $g<\kappa$. When
the two modes without the mode coupling in the WGMC are coupled to the DQE,
the photon antibunching is observed around $\Delta=0$. Such antibunching in
the weak DQE-field coupling regime results from an interference between
the coherent light transmitted through the resonant cavities and the
super-Poissonian light generated by photon-induced tunneling~\cite{45}. Thus a
balance between the decay rate $\kappa$ of the cavity modes and the
DQE-field coupling $g$ is required. When the DQE-field coupling is not
in the strong-coupling regime, the antibunching effect is weak, e.g., the
minimum value of $g_{\mathrm{CCW}}^{\mathrm{(2)}}(0)$ is about $0.47$ at
$\Delta=0$ for $\kappa=2g$, as shown in Fig.~\ref{Figure3}. However, when the
mode coupling $J$ is introduced, an obvious minimum and valley appear around
$\Delta=0$, e.g., the minimum value $g_{\mathrm{CCW}}^{\mathrm{(2)}}(0)$
reaches $0.006$ at $\Delta=0$, in contrast to the case when there is no mode
coupling. This strong antibunching is due to the mode coupling, which will be
discussed in the following paragraphs.

\begin{figure}[ptb]
\includegraphics[bb=93 219 535 640,  width=3.7cm, clip]{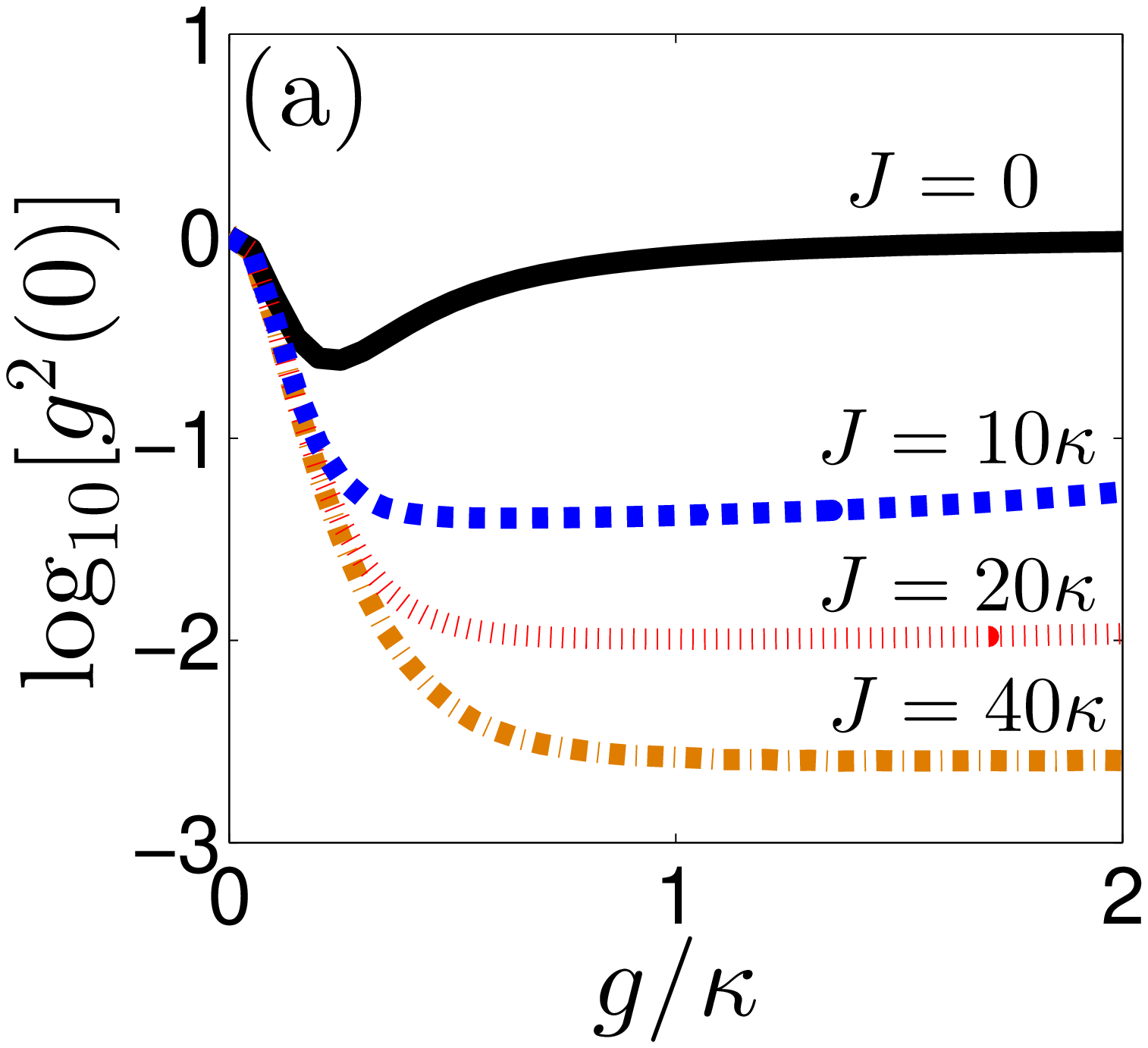}
\includegraphics[bb=23 205 539 636,  width=4cm, clip]{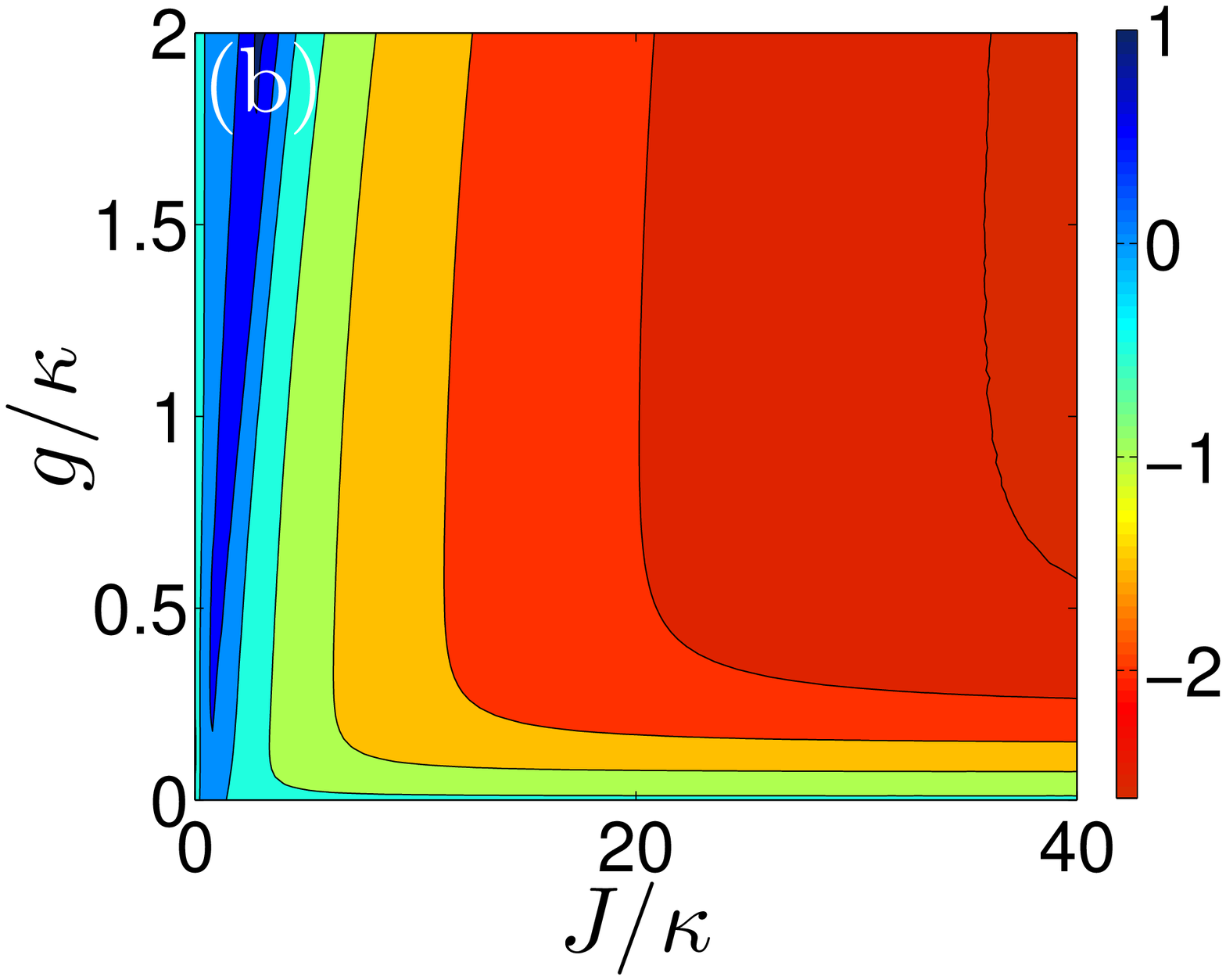}
\caption{(Color online) (a) The logarithmic of second-order correlation functions
$g_{\mathrm{a}}^{\mathrm{(2)}}(0)$ about the CCW-propagating intercavity field,
versus DQE-field coupling strength parameter $g/\kappa$, are calculated
and shown with different mode-coupling strengths: black-solid line for $J=0$;
blue-dashed line for $J=10\kappa$; red-dotted line for $J=20\kappa$, and
yellow-dot-dashed line for $J=40\kappa$. (b) The second-order correlation
function $g_{\mathrm{a}}^{\mathrm{(2)}}(0)$ of the CCW-propagating intercavity
field is plotted as functions of both the mode-coupling strengths $J/\kappa$ and
the DQE-field coupling strength parameter $g/\kappa$. The other system
parameters for this simulation are: $\kappa=40\gamma_{\mathrm{a}}$, $\Delta
=0$, $\varepsilon=\gamma_{\mathrm{a}}$, and $\gamma_{\mathrm{a}}=1$ MHz.}%
\label{Figure4}%
\end{figure}

To clearly show how the DQE-field coupling strength $g$ affects the
antibunching for different mode-coupling strengths $J$, including $J=0$,
$g_{\mathrm{CCW}}^{\mathrm{(2)}}(0)$ versus $g$ is plotted in
Fig.~\ref{Figure4}(a) for different values of $J$. As shown in
Fig.~\ref{Figure4}(a), for no mode-coupling, i.e., $J=0$, we find: (i) a
minimum value of $g_{\mathrm{CCW}}^{\mathrm{(2)}}(0)=0.25$ around of
$g/\kappa=0.2$, which is in the weak DQE-field coupling regime; (ii) the
value of $g_{\mathrm{CCW}}^{\mathrm{(2)}}(0)$ increases with increasing
$g/\kappa$ further; (iii) the value of $g_{\mathrm{CCW}}^{\mathrm{(2)}}(0)$ approaches
one when $g/\kappa>0.8$ and the photon antibunching gradually disappears;
(iv) the value of $g_{\mathrm{CCW}}^{\mathrm{(2)}}(0)$ equals to one and the
photon antibunching does not occur when $g/\kappa>1$, which is in the strong
DQE-field coupling regime. This is counterintuitive with the fact that the
better photon antibunching (i.e., the smaller value of $g_{\mathrm{CCW}%
}^{\mathrm{(2)}}(0)$) corresponds to a larger $g$. The physical mechanism for
this effect has been explained in Ref.~\cite{45}. Therefore, a balance between
$g$ and $\kappa$ is required, which might not be easy to achieve experimentally.
For example, photon antibunching can occur only in a narrow region around the
dip, e.g., $g/\kappa=0.2$. However, as shown in Fig.~\ref{Figure4}(a), when
mode-coupling is introduced, a better photon antibunching is achieved both in
the weak and strong DQE-field coupling regimes. Figure~\ref{Figure4}(a)
clearly shows that the better photon antibunching (smaller value of
$g_{\mathrm{CCW}}^{\mathrm{(2)}}(0)$) corresponds to the stronger
mode-coupling strength in both the weak and strong DQE-field coupling regimes. The
variations of $g_{\mathrm{CCW}}^{\mathrm{(2)}}(0)$ with $g/\kappa$ are similar
for different values of $J$.

We now study how the statistical properties of the cavity fields are affected
by the mode-coupling strength $J$ and the DQE-field coupling strength $g$
by plotting $g_{\mathrm{CCW}}^{\mathrm{(2)}}(0)$ in Fig.~\ref{Figure4}(b). We
find that a strong photon antibunching can be achieved both in the
DQE-field weak-coupling and strong-coupling for the strong mode-coupling, e.g.,
$J/\kappa>5$. For a given $g$, the larger $J$ corresponds to the smaller value
of $g_{\mathrm{CCW}}^{\mathrm{(2)}}(0)$, and thus a better photon antibunching
is achieved. Similarly, for a nonzero mode-coupling strength $J$, the larger
$g$ corresponds to the smaller value of $g_{\mathrm{CCW}}^{\mathrm{(2)}}(0)$,
and thus a better photon antibunching is achieved. From Fig.~\ref{Figure4}, we
also find that a nonzero mode-coupling strength $J$ makes the balance between
$g$ and $\kappa$  unnecessary, to obtain a strong photon
antibunching. This will be further discussed below.

\begin{figure}[ptb]
\includegraphics[bb=110 0 786 550,  width=8cm, clip]{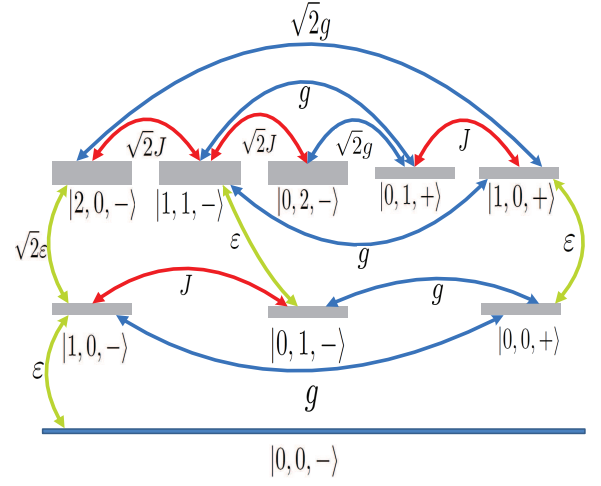}
\caption{(Coloronline) Schematic diagram of the energy levels of the WGMC-DQE system corresponding to
Fock states up to two-photons. The first photon and second photon indexes in
the ket correspond to the photon number in the CCW propagating mode and the CW
propagating mode, respectively. The index $+$ ($-$) indicate the excited state
(ground state) of the DQE. The long blue solid line at the bottom denotes the
ground state of the whole system. The narrower short gray lines indicate
single-photon states with decay rate $\kappa$, and the wider short gray lines
at the top represent two-photon states with decay rate $2\kappa$. Here we have
neglected the decay of the DQE due to the fact that $\kappa\gg\gamma_{a}$. The
four green double-arrows represent the driving optical pump for the CCW propagating
mode. The six blue double-arrows represent different energy level transitions
induced by the DQE-field coupling. The four red duble arrows represent
different energy level transitions induced by the CCW-CW mode coupling.}%
\label{figure6}%
\end{figure}

\subsection{Analytical solutions}

To better understand numerical calculations, we now study an analytical solution and compare it with numerical results. We assume that the driving field is very weak so that the total excitation number
of the system is no more than two~\cite{28,29,30,31,Ludwig-PRL2012,32,33}. In this
case, we use the ansatz%
\begin{align}
\left\vert \varphi(t)\right\rangle  &  =C_{0,0,-}\left\vert 0,0,-\right\rangle
+C_{1,0,-}\left\vert 1,0,-\right\rangle +C_{0,1,-}\left\vert
0,1,-\right\rangle \nonumber\label{eq:22}\\
&  +C_{0,0,+}\left\vert 0,0,+\right\rangle +C_{2,0,-}\left\vert
2,0,-\right\rangle +C_{0,2,-}\left\vert 0,2,-\right\rangle \nonumber\\
&  +C_{1,1,-}\left\vert 1,1,-\right\rangle +C_{1,0,+}\left\vert
1,0,+\right\rangle +C_{0,1,+}\left\vert 0,1,+\right\rangle ,
\end{align}
with $C_{i,j,\pm} \equiv C_{i,j,\pm}(t) $ , to calculate the steady state of the system. In the weak-driving limit, we have the relation%
\begin{align}
C_{0,0,-} &  \gg C_{1,0,-},C_{0,1,-},C_{0,0,+}\nonumber\label{eq:23}\\
&  \gg C_{2,0,-},C_{0,2,-},C_{1,1,-},C_{1,0,+},C_{0,1,+}.
\end{align}
Using the relation in Eq.~(\ref{eq:23}) and combining with Eq.~(\ref{Master}),
we have%
\begin{equation}
g_{\mathrm{CCW}}^{\mathrm{(2)}}(0)=\frac{2\left\vert C_{2,0,-}\right\vert
^{2}}{\left\vert C_{1,0,-}\right\vert ^{4}}.
\end{equation}
The coefficients $C_{2,0,-}$ and $C_{1,0,-}$ in the steady state can be
obtained via the Schr\"{o}dinger equation%
\begin{equation}
i\hbar\frac{\partial\left\vert \varphi\right\rangle }{\partial t}=\tilde
{H}\left\vert \varphi\right\rangle ,\label{xuedinge}%
\end{equation}
and $\partial\left\vert \varphi\right\rangle /\partial t=0$. The effective
non-hermitian Hamiltonian $\tilde{H}$ in Eq.~(\ref{xuedinge}) is given
by
\begin{align}
\tilde{H} &  =\hbar\left(  \Delta-i\frac{\kappa}{2}\right)  \left(  a^{\dag
}a+b^{\dag}b\right)  +\hbar J\left(  a^{\dag}b+b^{\dag}a\right)
\label{hamiton}\\
&  +\hbar\left(  \Delta-i\frac{\gamma_{\mathrm{a}}}{2}\right)  \sigma
_{z}+\hbar\left[  g(a^{\dag}+b^{\dag})\sigma_{-}+\varepsilon a^{\dag
}+\text{H.c.}\right]  ,\nonumber
\end{align}
in the zero temperature approximation. Here, the dephasing of the DQE has been neglected.

Then, replacing $\tilde{H}$ in Eq.~(\ref{xuedinge}) by the expression in
Eq.~(\ref{hamiton}), we can obtain a set of linear equations. the detailed derivation can be found in Appendix A. By solving these
linear equations, we can obtain the coefficients, e.g. $C_{2,0,-}$ and
$C_{1,0,-}$, in Eq.~(\ref{eq:22}), then $g_{\mathrm{CCW}}^{\mathrm{(2)}}(0)$
is further calculated as%
\begin{equation}
g_{\mathrm{CCW}}^{(2)}(0))=\frac{2\left\vert C_{2,0,-}\right\vert ^{2}%
}{\left\vert C_{1,0,-}\right\vert ^{4}}=\frac{\left\vert A_{1}\right\vert
^{2}\left\vert A_{2}\right\vert ^{2}}{\left\vert A_{3}\right\vert
^{2}\left\vert A_{4}\right\vert ^{4}},\label{JieXi}%
\end{equation}
with%
\begin{align}
A_{1} &  =\Delta_{\mathrm{p}}[\Delta_{\mathrm{p}}^{2}\Delta_{\mathrm{d}%
}+(\Delta_{\mathrm{d}}\Delta_{\mathrm{p}}-2g^{2})(\Delta_{\mathrm{d}%
}+J)]+g^{4},\\
A_{2} &  =\Delta_{\mathrm{d}}(J+\Delta_{\mathrm{p}})-2g^{2},\\
A_{3} &  =(J+\Delta_{\mathrm{p}})(J+\Delta_{\mathrm{p}}+\Delta_{\mathrm{d}%
})-2g^{2},\\
A_{4} &  =\Delta_{\mathrm{p}}\Delta_{\mathrm{d}}-g^{2}.
\end{align}
Here, we define two detunnings by including the dissipation rates%
\begin{align}
\Delta_{\mathrm{p}} &  =\Delta-i\frac{\kappa}{2},\\
\Delta_{\mathrm{d}} &  =\Delta-i\frac{\gamma_{\mathrm{a}}}{2}.
\end{align}

For comparison with the numerical calculations using the master equation, $g_{\mathrm{CCW}%
}^{\mathrm{(2)}}(0)$, calculated by analytical solution Eq.~(\ref{JieXi}),
versus $\Delta/\kappa$, is also shown in Fig.~\ref{Figure3}. It is clear that
the analytical solutions agree well with the numerical
calculations.

\begin{figure}[ptb]
\includegraphics[bb=9 147 571 653,  width=8cm, clip]{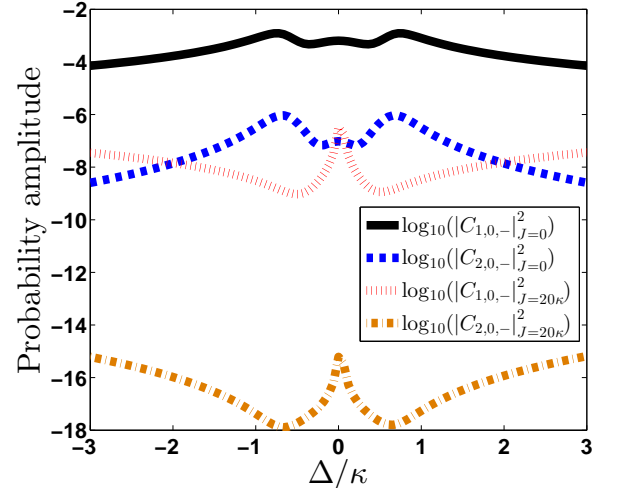}
\caption{(Color online) The probability amplitude of the WGMC-DQE system with
one-photon and two-photons versus the detunning parameter $\Delta/\kappa$ are
calculated and shown with different mode-coupling strengths: the black-solid curve
for the one-photon probability amplitude $\left\vert C_{1,0,-}\right\vert ^{2}$
and blue-dashed curve for the two-photon probability amplitude $\left\vert
C_{2,0,-}\right\vert ^{2}$ with $J=0$; the red-dotted curve for the one-photon
probability amplitude $\left\vert C_{1,0,-}\right\vert ^{2}$ and the yellow dot
dash curve for the two-photon probability amplitude $\left\vert C_{2,0,-}%
\right\vert ^{2}$\ with $J=20\kappa$. The other system parameters for this
simulation are: $\kappa=40\gamma_{\mathrm{a}},$ $g=20\gamma_{\mathrm{a}}$,
$\varepsilon=\gamma_{\mathrm{a}}$, and $\gamma_{\mathrm{a}}=1$ MHz.}%
\label{Figure7}%
\end{figure}

\subsection{Physical mechanism and comparisons between with and without mode
coupling}

Using analytical solutions, let us further understand the physical
mechanism of the mode-coupling-enhanced photon antibunching. In
Fig.~\ref{figure6}, we schematically show the energy level structure of the
system up to two-particle excitations. In this case, nine states are involved,
they are the ground state $\left\vert 0,0,0\right\rangle $, three one-particle
excitation states $\left\vert 1,0,-\right\rangle $, $\left\vert
0,1,-\right\rangle $, and $\left\vert 0,0,+\right\rangle $, and five two-particle
excitation states $\left\vert 2,0,-\right\rangle $, $\left\vert
0,2,-\right\rangle $, $\left\vert 1,1,-\right\rangle $, $\left\vert
1,0,+\right\rangle $, $\left\vert 0,1,+\right\rangle $. The connections
between different states, due to the driving field $\varepsilon$ and the
couplings, characterized by $J$ and $g$, are also schematically shown in
Fig.~\ref{figure6}.

We first consider the single-particle excitation. In this case, the CCW mode
is pumped to the single-photon state and the system is in the state
$\left\vert 1,0,-\right\rangle $. When there is no mode coupling, i.e., $J=0$,
the probability amplitude $\left\vert C_{1,0,-}\right\vert ^{2}$, which
represents the probability to find the system in the state $\left\vert
1,0,-\right\rangle $, is only determined by the decay rate $\kappa$ of the CCW
mode and DQE-field coupling strength $g$. In this case and as shown in
Fig.~\ref{figure6}, there are two transition paths from the state $\left\vert
1,0,-\right\rangle $ to the state $\left\vert 0,0,+\right\rangle $ or the
state $|0,0,-\rangle$. However, when the mode coupling is introduced by the
scatter, $\left\vert C_{1,0,-}\right\vert ^{2}$ depends not only on $\kappa$
and $g$ but also on the mode-coupling strength $J$. Thus an additional transition
path between the state $\left\vert 1,0,-\right\rangle $ and the state
$\left\vert 0,1,-\right\rangle $ is introduced by the mode coupling, which may
reduce the probability of the system at the state $\left\vert
1,0,-\right\rangle $.

When the CCW mode is further pumped to the two-photon state $\left\vert
2,0,-\right\rangle $. The additional transition path between the state
$\left\vert 2,0,-\right\rangle $ and the state $\left\vert 1,1,-\right\rangle
$ is introduced due to the mode coupling, and this may also reduce the probability of
the system in the state $\left\vert 2,0,-\right\rangle $. Therefore, when
mode coupling $J$ is introduced, both the single-photon probability
$\left\vert C_{1,0,-}\right\vert ^{2}$ and two-photon probability $\left\vert
C_{2,0,-}\right\vert ^{2}$ might be reduced, simultaneously. Considering
Eq.~(\ref{JieXi}), we now ask a natural question: which probability is larger,
because only the ratio between $\left\vert C_{2,0,-}\right\vert ^{2}$ and
$\left\vert C_{1,0,-}\right\vert ^{4}$ makes sense.

The probability amplitudes $\left\vert C_{1,0,-}\right\vert ^{2}$ and
$\left\vert C_{2,0,-}\right\vert ^{2}$ versus detunning $\Delta/\kappa$ are
plotted in Fig.~\ref{Figure7} for $J=0$ and $J=20\kappa$. We analyze
Fig.~\ref{Figure7} in two parameter regimes, i.e., $|\Delta|/\kappa<0.6$ and
$|\Delta|/\kappa>0.6$. In the first parameter regime and around $\Delta=0$, it
is clear that both $\left\vert C_{1,0,-}\right\vert ^{2}$ and $\left\vert
C_{2,0,-}\right\vert ^{2}$ are reduced when the mode coupling is introduced.
They are changed as follows%
\begin{equation}
\left\vert C_{2,0,-}\right\vert _{J=20\kappa}^{2}=10^{-8}\left\vert
C_{2,0,-}\right\vert _{J=0}^{2},%
\end{equation}
and%
\begin{equation}
\left\vert C_{1,0,-}\right\vert _{J=20\kappa}^{2}=10^{-3}\left\vert
C_{1,0,-}\right\vert _{J=0}^{2}.
\end{equation}
Considering the expression of $g_{\mathrm{CCW}}^{(2)}(0)$, we find that
$g_{\mathrm{CCW}}^{\mathrm{(2)}}(0)$ is reduced by two orders of magnitude when
the mode coupling is introduced, i.e., stronger photon antibunching is
achieved. When $\left\vert \Delta\right\vert /\kappa>0.6$, both $\left\vert
C_{2,0,-}\right\vert ^{2}$ and $\left\vert C_{1,0,-}\right\vert ^{4}$ are
comparable for the case with mode coupling or without mode coupling, thus
$g_{\mathrm{CCW}}^{(2)}(0)$ are the same for the cases with and without mode
coupling in this parameter regime.

We can also obtain the value of $g_{\mathrm{CCW}}^{\mathrm{(2)}}(0)$ via
analytical solution in Eq.~(A9). When $J=0$, we have%
\begin{equation}
\left\vert C_{1,0,-}\right\vert _{J=0}^{2}=\frac{\varepsilon^{2}\left\vert
\Delta_{p}\Delta_{d}-g^{2}\right\vert ^{2}}{\left\vert \Delta_{p}\right\vert
^{2}\left\vert -2g^{2}+\Delta_{p}\Delta_{d}\right\vert ^{2}}.
\end{equation}
Around $\Delta=0$, the ratio of $\left\vert C_{1,0,-}\right\vert ^{2}$ between
$J=0$ and $J\neq0$ is given%
\begin{equation}
R_{1}=\frac{\left\vert C_{1,0,-}\right\vert _{J\neq0}^{2}}{\left\vert
C_{1,0,-}\right\vert _{J=0}^{2}}\approx\frac{\kappa^{2}}{4J^{2}},
\end{equation}
Similarly, we can have
\begin{equation}
R_{2}=\frac{\left\vert C_{2,0,-}\right\vert _{J\neq0}^{2}}{\left\vert
C_{2,0,-}\right\vert _{J=0}^{2}}\approx\frac{\kappa^{6}}{4J^{6}}.
\end{equation}
Then, we can obtain $g_{\mathrm{CCW}}^{\mathrm{(2)}}(0)$
as%
\begin{equation}
\frac{g_{\mathrm{CCW}}^{2}(0)_{J\neq0}}{g_{\mathrm{CCW}}^{2}(0)_{J=0}}%
=\frac{R_{2}}{R_{1}^{2}}=\frac{4\kappa^{2}}{J^{2}}.
\end{equation}
When $J=20\kappa$, we find that $g_{\mathrm{CCW}}^{\mathrm{(2)}}%
(0)\approx10^{-2}$ in the range near $\Delta=0$, which is the same as
numerical calculations.

\begin{figure}[ptb]
\includegraphics[bb=10 173 577 664,  width=4cm, clip]{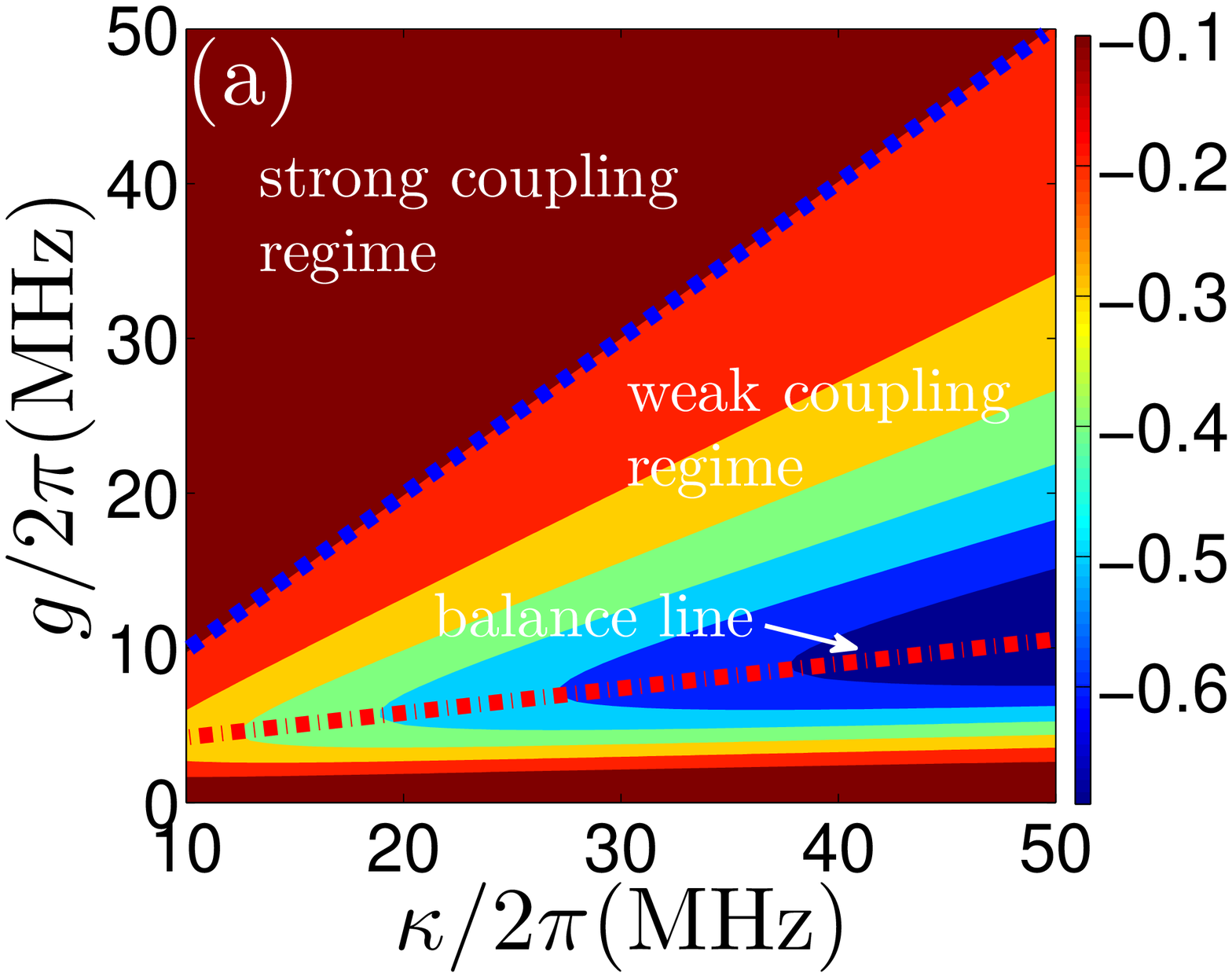}
\includegraphics[bb=10 173 577 664,  width=4cm, clip]{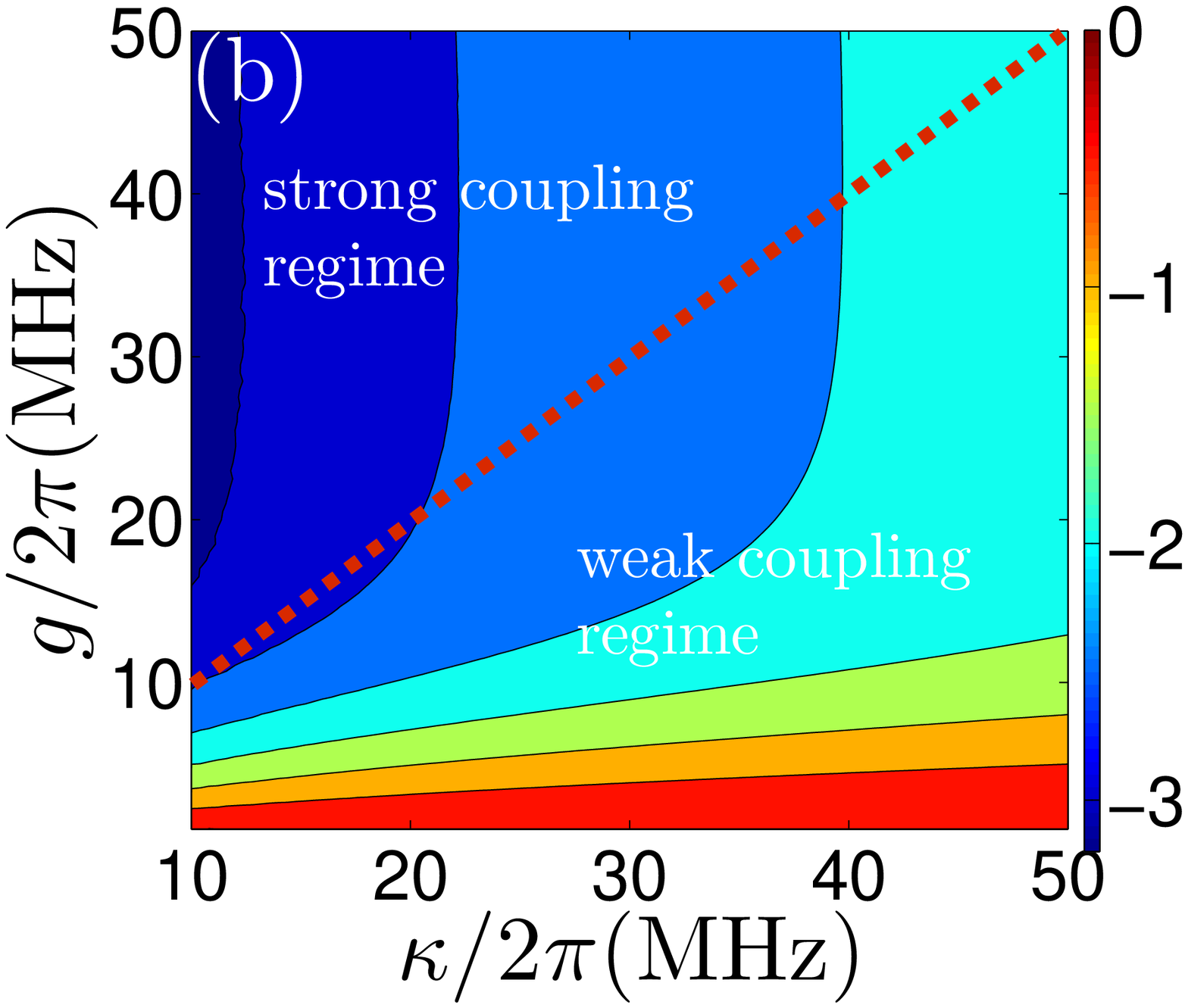}
\caption{(Color online) Logarithmic plot of second-order correlation functions $g_{\mathrm{CCW}}%
^{\mathrm{(2)}}(0)$ of the CCW-propagating intercavity field as
functions of both cavity field decay rate $\kappa$ and DQE-field coupling
strength parameter $g$ without optical mode coupling in (a) and with optical
mode coupling rate $J=800$ MHz in (b). The balance line between parameter g and $\kappa$ is shown as a dash-dotted line in (a). The strong and weak DQE-field coupling regime are labeled up and below the dash line respectively. The other system parameters are:
$\Delta=0$, $\varepsilon=\gamma_{\mathrm{a}}$, $\gamma_{\mathrm{a}}=1$ MHz.}%
\label{figure8}%
\end{figure}

According to the above discussions, we conclude that the mode
coupling results in the strong photon antibunching in the range near
$\Delta=0$. To clearly show this, $g_{\mathrm{CCW}}^{\mathrm{(2)}}(0)$ is
further plotted as a function of $\kappa$ and $g$ with $J=0$ in
Fig.~\ref{figure8}(a) and $J=800$ MHz in Fig.~\ref{figure8}(b). In contrast to
the case of DQE-bimodal system without mode coupling~\cite{23}, the strong
photon antibunching appears in the larger parameter range of the DQE-field
coupling strength $g$ when the mode coupling is introduced. Comparing this with no
photon antibunching, i.e., $g_{\mathrm{CCW}}^{\mathrm{(2)}}(0)\sim1$, in the
strong DQE-field coupling regime when there is no the mode-coupling, we
find that a strong photon antibunching, i.e., $g_{\mathrm{CCW}}^{\mathrm{(2)}%
}(0)\sim10^{-2}$, appears in the strong DQE-field coupling regime when the
mode coupling is introduced. We also find that a balance between $\kappa$ and
$g$ is necessary to obtain the stronger photon antibunching in the weak
DQE-field coupling regime when $J=0$. However, when the mode coupling $J$
is introduced, such balance is not required.

\section{Mode coupling enhanced robustness to pure dephasing}

It is well known that dephasing with the rate $\gamma_{\mathrm{p}}$ cannot
be neglected in some particular DQEs, e.g., the dephasing due to
electron-phonon coupling play a very important role in quantum dot systems. Thus,
we now study the pure dephasing effect and show how it affects the statistical
properties of CCW propagating optical mode in the weak DQE-field coupling regime.

\begin{figure}[ptb]
\includegraphics[bb=10 173 577 664,  width=4cm, clip]{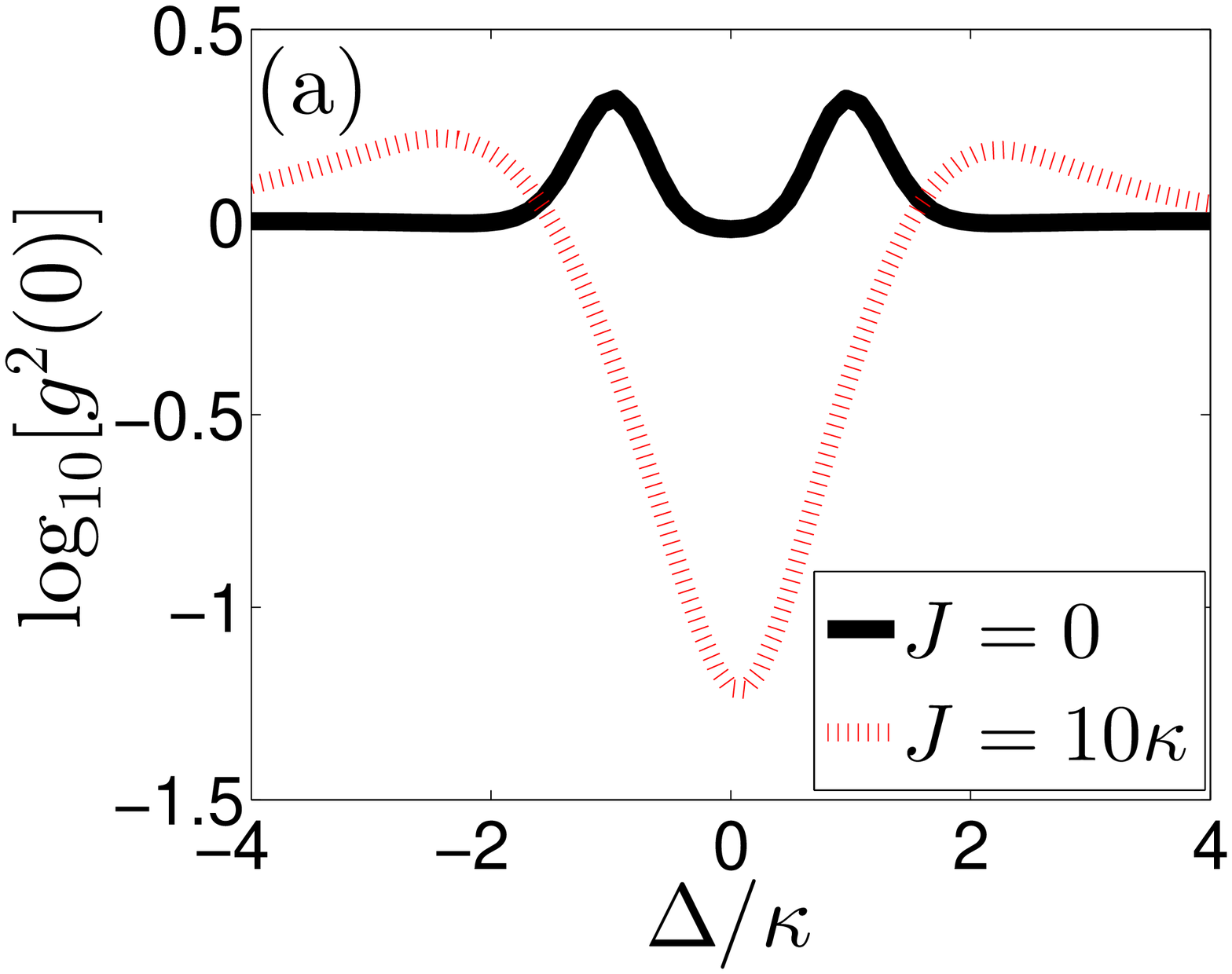}
\includegraphics[bb=10 173 577 664,  width=4cm, clip]{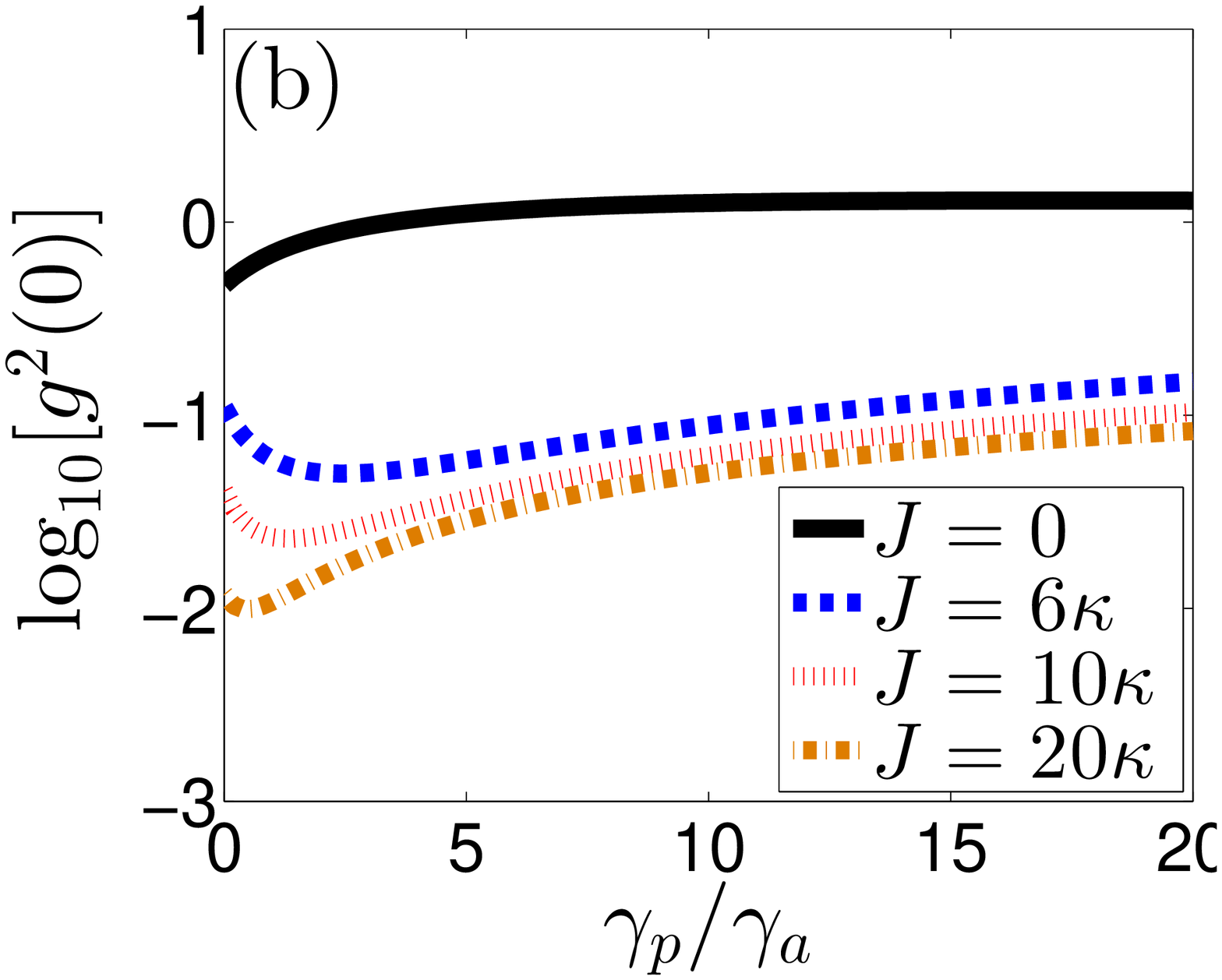}
\caption{(Color online) (a) The second-order correlation functions
$g_{\mathrm{CCW}}^{\mathrm{(2)}}(0)$ of the CCW-propagating intercavity fields are
plotted as functions of the cavity-field detuning $\Delta/\kappa$, with
$\gamma_{\mathrm{p}}=3\gamma_{\mathrm{a}}$. The black-solid curve is for the
cavity mode without coupling ($J=0$) and red-dash-dotted curve is for strong
mode-coupling ($J=10\kappa$). (b) The $g_{\mathrm{CCW}}^{\mathrm{(2)}}(0)$ versus
the dephasing rate $\gamma_{p}/\gamma_{a}$ with different mode-coupling strength:
$J=0$ (black-solid curve), $J=6$ (blue-dash curve), $J=10\kappa$
(red-dotted curve) and $J=20\kappa$ (yellow-dotted-dash curve) at $\Delta=0$.
The other system parameters for these calculations are: $\kappa=40\gamma
_{\mathrm{a}}$, $g=20\gamma_{\mathrm{a}}$, $\varepsilon=\gamma_{\mathrm{a}}$,
$\gamma_{\mathrm{a}}=1$ MHz.}%
\label{Figure9}%
\end{figure}

We first study $g_{\mathrm{CCW}}^{\mathrm{(2)}}(0)$ as a function of the
detuning $\Delta$. Figure~\ref{Figure3} has shown that the photon antibunching
can be achieved around $\Delta=0$ for $J=0$ when there is no dephasing.
However, as shown in Fig.~\ref{Figure9}(a), when a pure-dephasing, e.g.,
$\gamma_{\mathrm{p}}=3\gamma_{\mathrm{a}}$, is introduced, the photon
antibunching effect disappears with all the values of $g_{\mathrm{CCW}%
}^{\mathrm{(2)}}(0)>1$ for $J=0$. Thus, if there is no the mode
coupling~\cite{23}, the photon antibunching is not easy to be achieved when
there is the dephasing. However, when the mode coupling is introduced, e.g.,
$J=10\kappa$, as shown in Fig.~\ref{Figure9}(a), a strong photon antibunching
can be achieved even when there is the dephasing. The minimum value of
$g_{\mathrm{CCW}}^{\mathrm{(2)}}(0)$ corresponding to $J=10\kappa$ is 0.06 at
the point $\Delta=0$ in contrast to $g_{\mathrm{CCW}}^{\mathrm{(2)}}(0)=1$ for
$J=0$.

We further study how the dephasing affects $g_{\mathrm{CCW}}^{\mathrm{(2)}%
}(0)$ in the weak DQE-field coupling regime in different mode-coupling
strengths. Figure~\ref{Figure9}(b) shows that $g_{\mathrm{CCW}}^{\mathrm{(2)}%
}(0)$ increases with increasing $\gamma_{\mathrm{p}}$ and quickly
approaches to one for $J=0$. However, when mode coupling is
introduced (as shown in Fig.~\ref{Figure9}(b) for three different
mode-coupling strengths, e.g., $J=6\kappa$, $J=10\kappa$, and $J=20\kappa$),
the photon antibunching can still occur even with large dephasing.
Interestingly, there is a minimum value of $g_{\mathrm{CCW}}^{\mathrm{(2)}%
}(0)$ versus of $\gamma_{\mathrm{p}}$ for each mode-coupling strength, and
this minimum value represents the optimal case for photon antibunching. We
find that photon antibunching can still be achieved when $\gamma_{\mathrm{p}}$
becomes very large. This means that the system is robust to dephasing when
mode coupling is introduced. Figure~\ref{Figure9}(b) also shows that larger
value of the mode-coupling strength $J$ provides a better photon antibunching
when the DQE has dephasing.

\begin{figure}[ptb]
\includegraphics[bb=10 173 577 664,  width=4.2cm, clip]{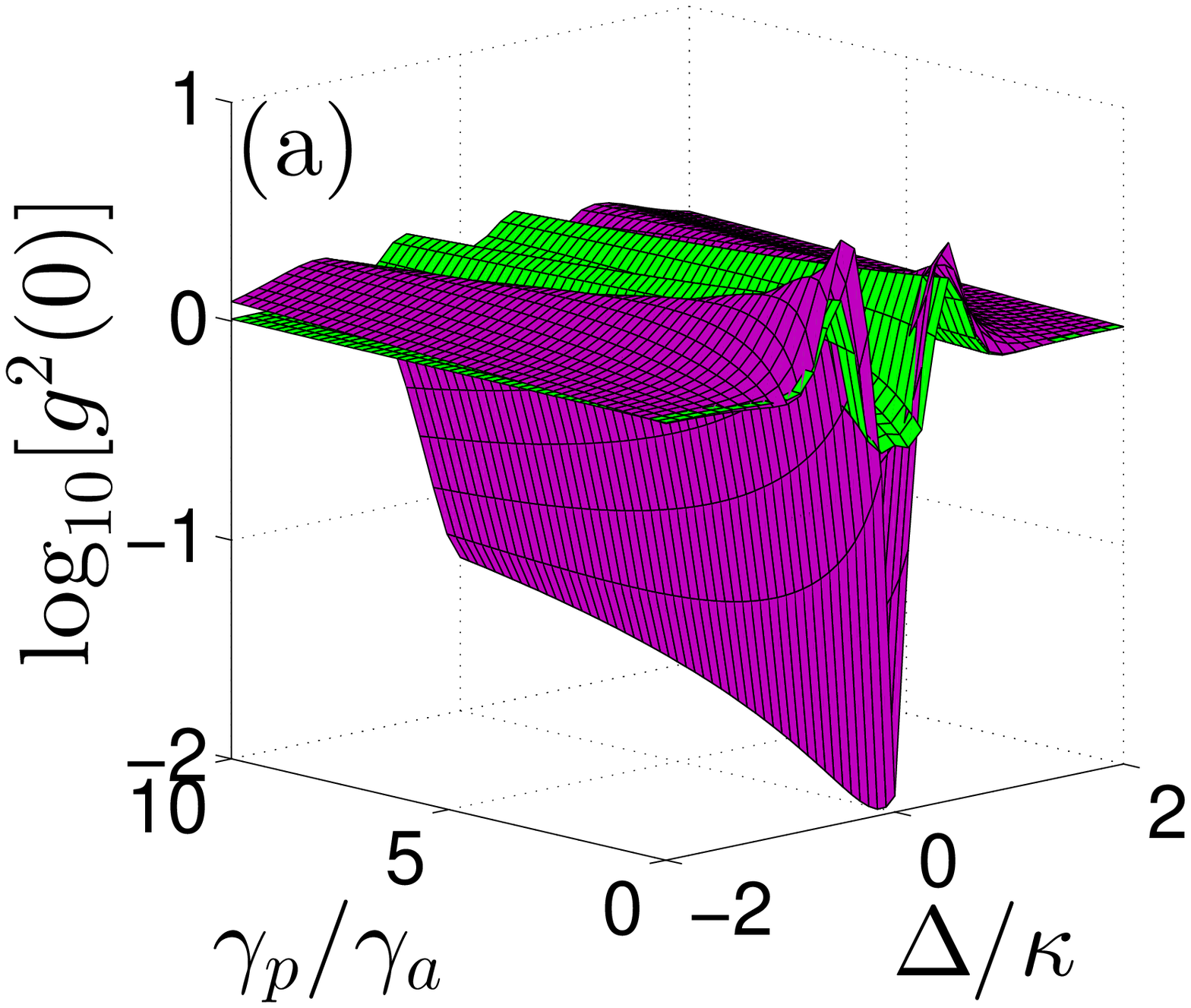}
\includegraphics[bb=10 173 577 664,  width=4.2cm, clip]{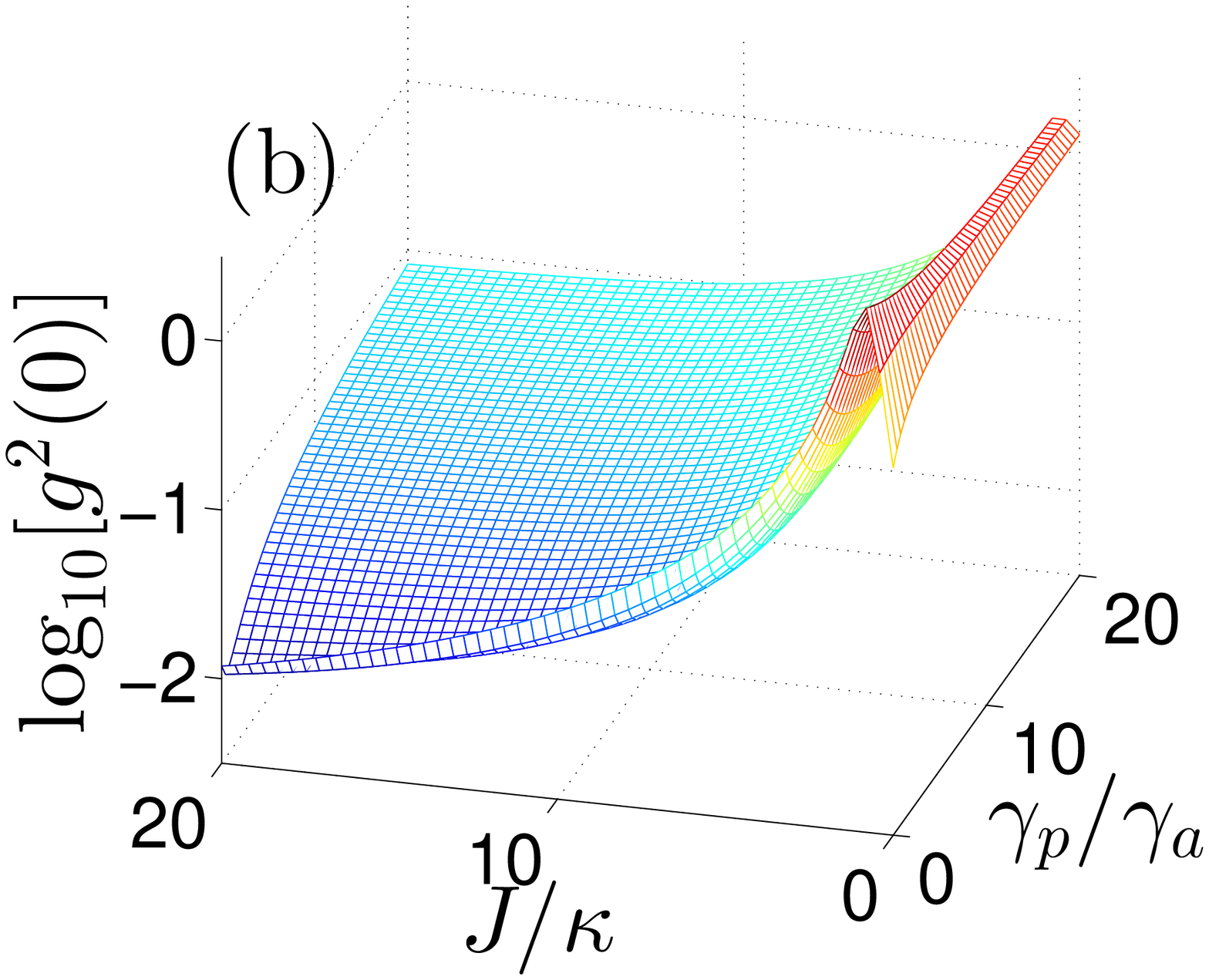}
\caption{(Color online) The second-order correlation functions
$g_{\mathrm{CCW}}^{\mathrm{(2)}}(0)$ of the CCW-propagating intercavity fields
are plotted in (a) as functions of both the cavity field detuning
$\Delta/\kappa$ and DQE dephasing rate $\gamma_{p}$, without mode
coupling $J=0$ (the cyan-curved surface), and with mode coupling
$J=20\kappa$ (the magenta-curved surface). (b) shows $g_{\mathrm{CCW}%
}^{\mathrm{(2)}}(0)$ versus both mode-coupling strength $J/\kappa$ and
DQE dephasing rate $\gamma_{p}$ at $\Delta=0$. The other system parameters
are: $\kappa=40\gamma_{\mathrm{a}}$, $g=20\gamma_{\mathrm{a}}$, $\varepsilon
=\gamma_{\mathrm{a}}$, and $\gamma_{\mathrm{a}}=1$ MHz.}%
\label{Figure10}%
\end{figure}

In Fig.~\ref{Figure10}(a), the $g_{\mathrm{CCW}}^{\mathrm{(2)}}(0)$ is
plotted as a function of $\Delta$ and $\gamma_{\mathrm{p}}$ with $J=0$ and
$J\neq0$, e.g., $J=20\kappa$. For $J=0$, shown as a cyan-curved surface, in
Fig.~\ref{Figure10}(a), $g_{\mathrm{CCW}}^{\mathrm{(2)}}(0)$ quickly
approaches to one around $\Delta=0$ with increasing $\gamma_{\mathrm{p}}%
$. This means that the antibunching is significantly affected by the
dephasing. However, for $J=20\kappa$, shown as a magenta-curved surface, in
Fig.~\ref{Figure10}(a), $g_{\mathrm{CCW}}^{\mathrm{(2)}}(0)$ is slowly
increased with increasing $\gamma_{\mathrm{p}}$ around $\Delta=0$, and remains smaller than
one up to $\gamma_{p}=10\,\gamma_{a}$. Thus the system for the photon
antibunching is very robust to the dephasing $\gamma_{\mathrm{p}}$ when the
mode coupling is introduced.

In Fig.~\ref{Figure10}(b), $g_{\mathrm{CCW}}^{\mathrm{(2)}}(0)$ is plotted as
a function of $J$ and $\gamma_{\mathrm{p}}$. Figure~\ref{Figure10}(b) shows
that the larger mode-coupling strength $J$ corresponds to the smaller value of
$g_{\mathrm{CCW}}^{\mathrm{(2)}}(0)$ for a given $\gamma_{p}$, and thus a better
photon antibunching. It is also obvious that all the values of
$g_{\mathrm{CCW}}^{\mathrm{(2)}}(0)$ are smaller than $0.1$ for all $\gamma
_{p}$ when $J/\kappa>4$. This reveals that strong photon antibunching can
occur even when a large dephasing $\gamma_{p}$ is introduced (e.g., for
$\gamma_{\mathrm{p}}/\gamma_{\mathrm{a}}=20$ and even larger). This is very
different from the case without the mode coupling, i.e., the value of
$g_{\mathrm{CCW}}^{\mathrm{(2)}}(0)$ approaches one even with a small
dephasing, as shown in Fig.~\ref{Figure9}(b) for $J=0$. Therefore, we conclude
that photon antibunching is robust to the phase dephasing of the DQE when
the mode coupling is introduced.

\section{Robustness to mode mismatch and coupling mismatch}

In previous sections, we mostly focused on the resonant interaction between
the DQE and cavity modes and also the homogeneous couplings between two
cavity modes and the DQE, i.e., $\omega=\omega_{a}$ and $g_{a}\equiv
g_{b}\equiv g$. Let us now analyze the effect of frequency mismatch and
coupling mismatch on photon antibunching.

\begin{figure}[ptb]
\includegraphics[bb=10 173 577 664,  width=4.2cm, clip]{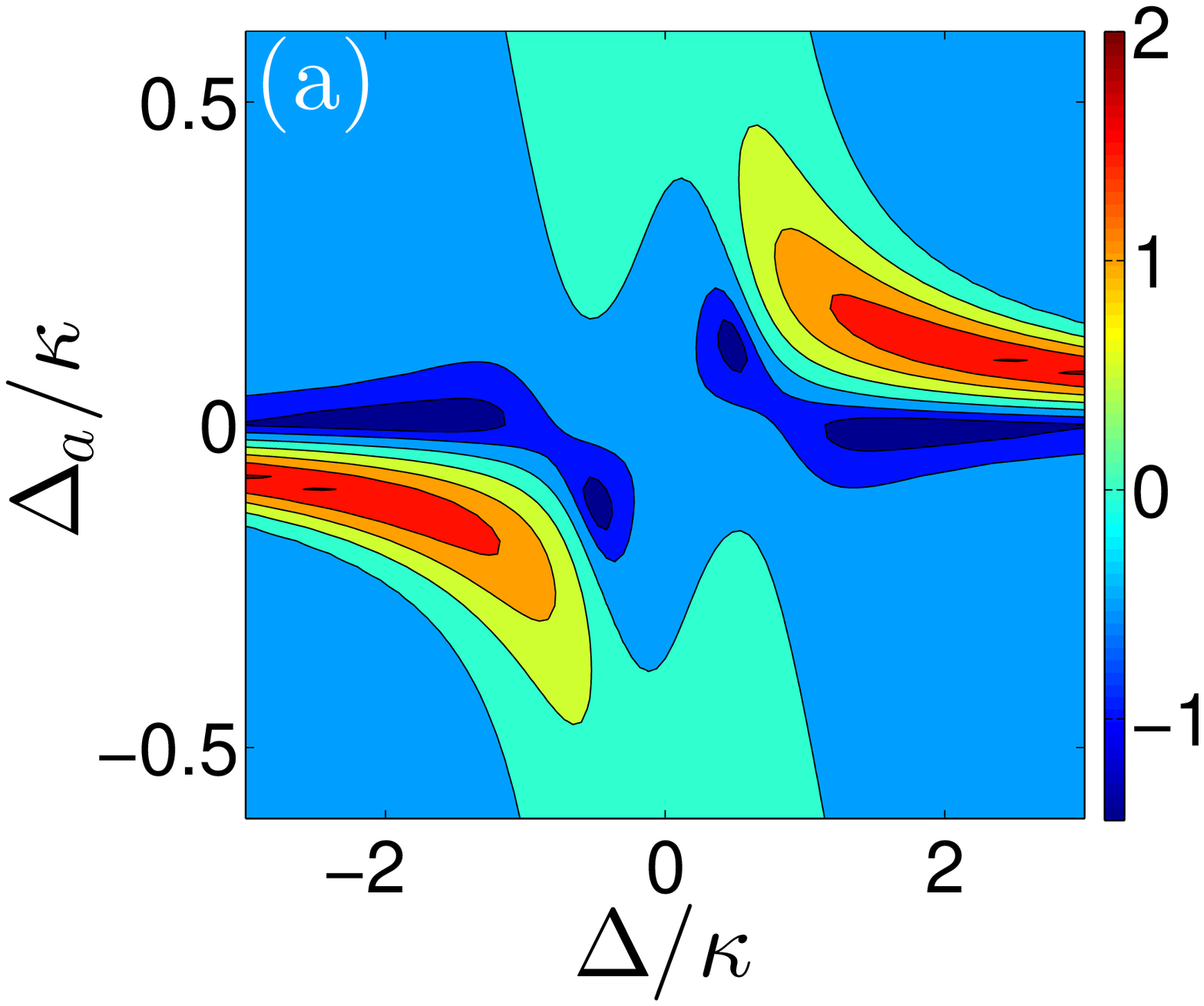}
\includegraphics[bb=10 173 577 664,  width=4.2cm, clip]{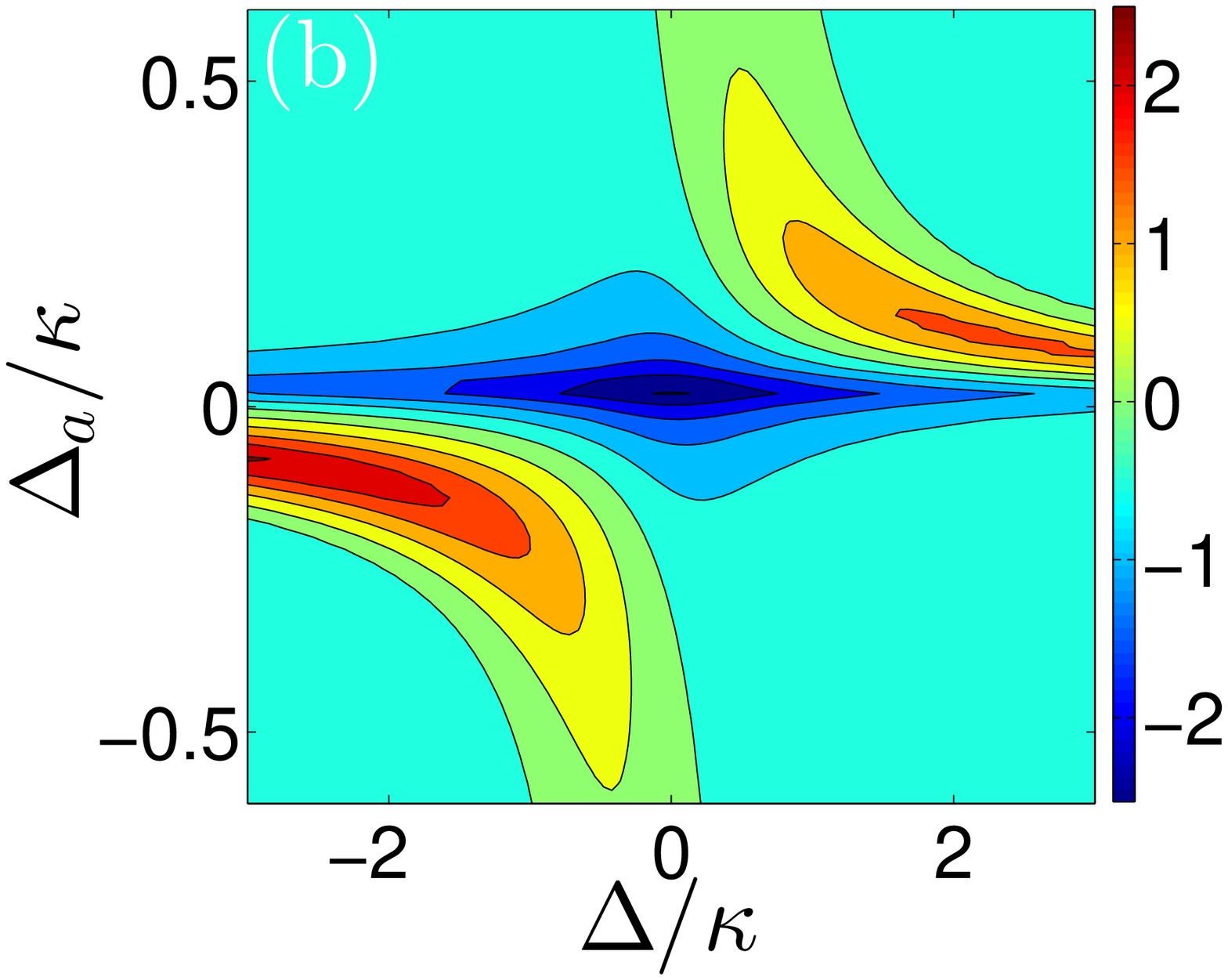}
\caption{(Color online) The second-order correlation functions
$g_{\mathrm{CCW}}^{\mathrm{(2)}}(0)$ of the CCW-propagating intercavity field
versus both the optical mode detuning $\Delta/\kappa$ and the DQE detuning
$\Delta_{a}/\kappa$, for (a) $J=0$ and (b) $J=20\kappa$. The other system
parameters are: $\kappa=40\gamma_{\mathrm{a}}$, $g=20\gamma_{\mathrm{a}}$,
$\varepsilon=\gamma_{\mathrm{a}}$, and $\gamma_{\mathrm{a}}=1$ MHz.}%
\label{Figure14}%
\end{figure}

We first study the effect of the mismatch between the frequencies $\omega$ and
$\omega_{\mathrm{a}}$ of the cavity modes and the DQE on the antibunching when mode coupling is introduced. In Fig.~\ref{Figure14}, for given $g_{\mathrm{a}}=g_{\mathrm{b}}=0.5\kappa$, the
second-order correlation function $g_{\mathrm{CCW}}^{\mathrm{(2)}}(0)$ is
plotted as a function of two detunings $\Delta=\omega-\omega_{d}$ and
$\Delta_{\mathrm{a}}=\omega_{a}-\omega_{d}$, for $J=0$ and also for $J\neq0$,
e.g., $J=20\kappa$. We find that $g_{\mathrm{CCW}}^{\mathrm{(2)}}(0)$ reaches
its minimum value when the driving field is resonant to the DQE,
e.g., $\Delta_{a}=0$. If we fix $\Delta_{a}\equiv0$, the effect of the
mismatch of the frequencies $\omega$ and $\omega_{\mathrm{a}}$ on the
antibunching can be clearly observed only by changing the detuning $\Delta$.
Compared with Fig.~\ref{Figure14}(a), in Fig.~\ref{Figure14}(b), we find that
increasing the mismatch $\left\vert \Delta\right\vert $, the value of
$g_{\mathrm{CCW}}^{\mathrm{(2)}}(0)$ increases very slowly.
The value of $g_{\mathrm{CCW}}^{\mathrm{(2)}}(0)$ increases to about $0.1$
when $\Delta$ is changed from zero to a finite value for $\Delta
_{\mathrm{a}}=0$. So, when mode coupling is introduced, the
photon antibunching can always be obtained in any detuning $\Delta$ when
the driving field is resonantly applied to the DQE, i.e., when $\Delta_{a}=0$.

\begin{figure}[ptb]
\includegraphics[bb=10 173 577 664,  width=6cm, clip]{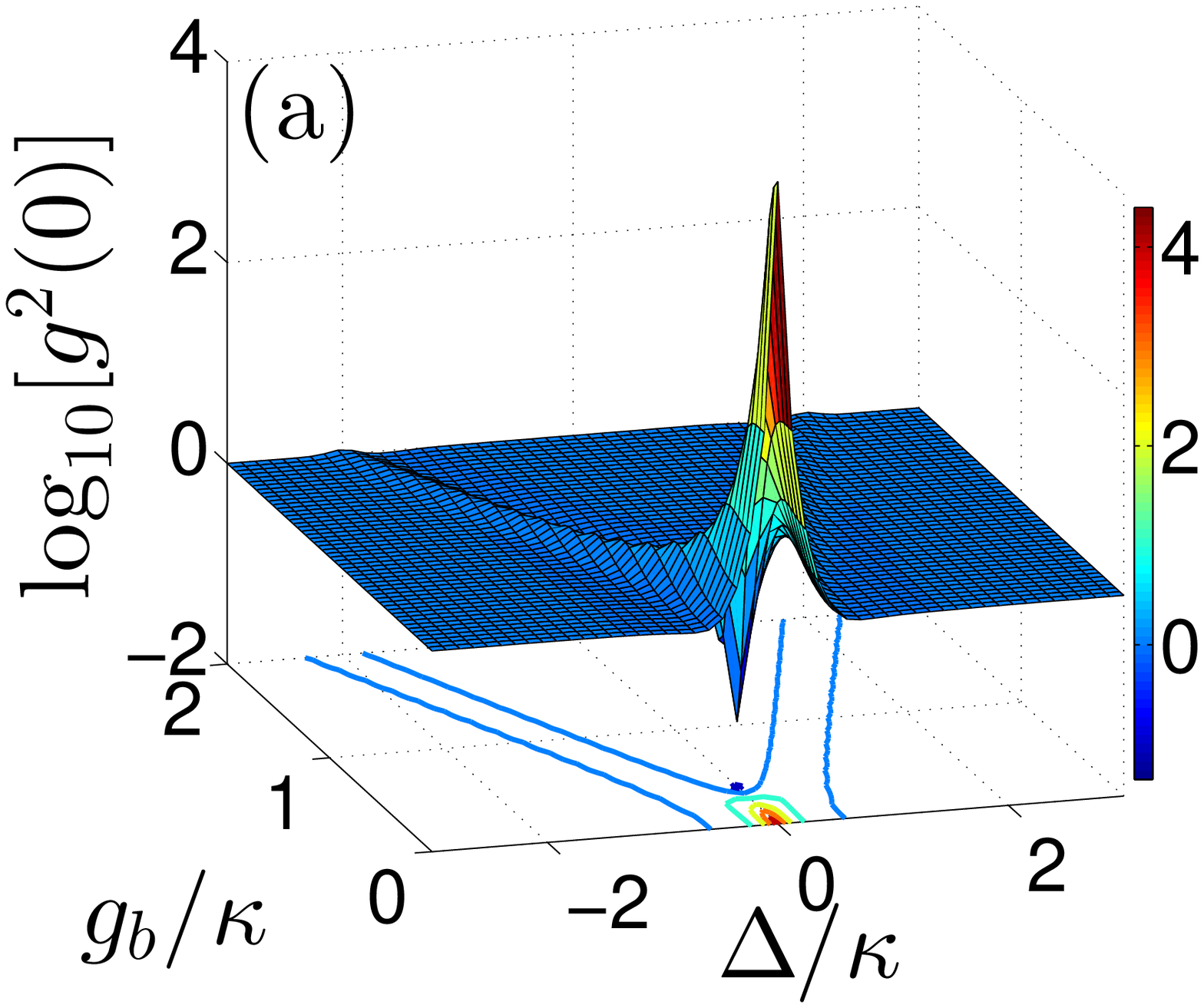}
\includegraphics[bb=10 173 577 664,  width=6cm, clip]{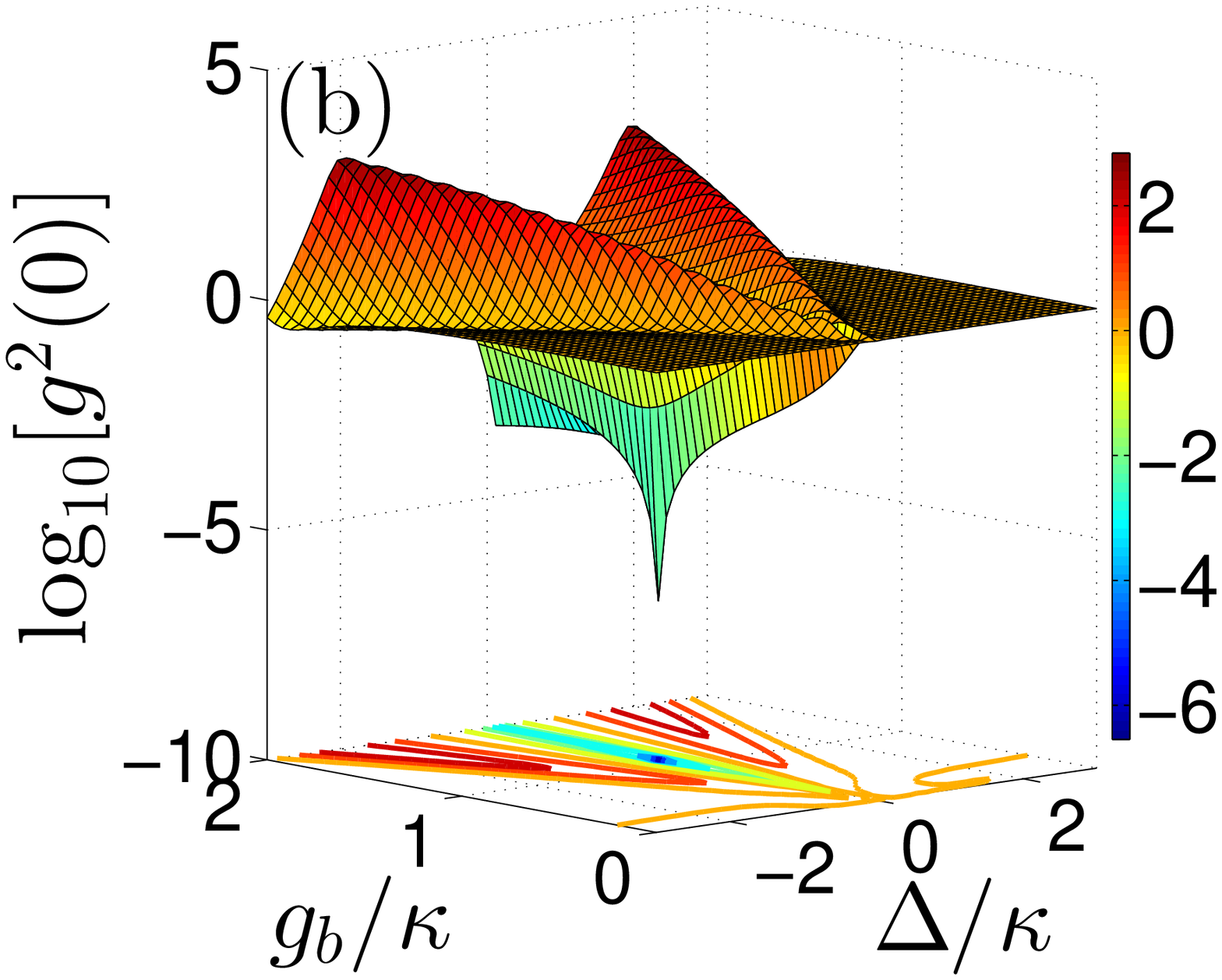}
\includegraphics[bb=10 173 577 664,  width=6cm, clip]{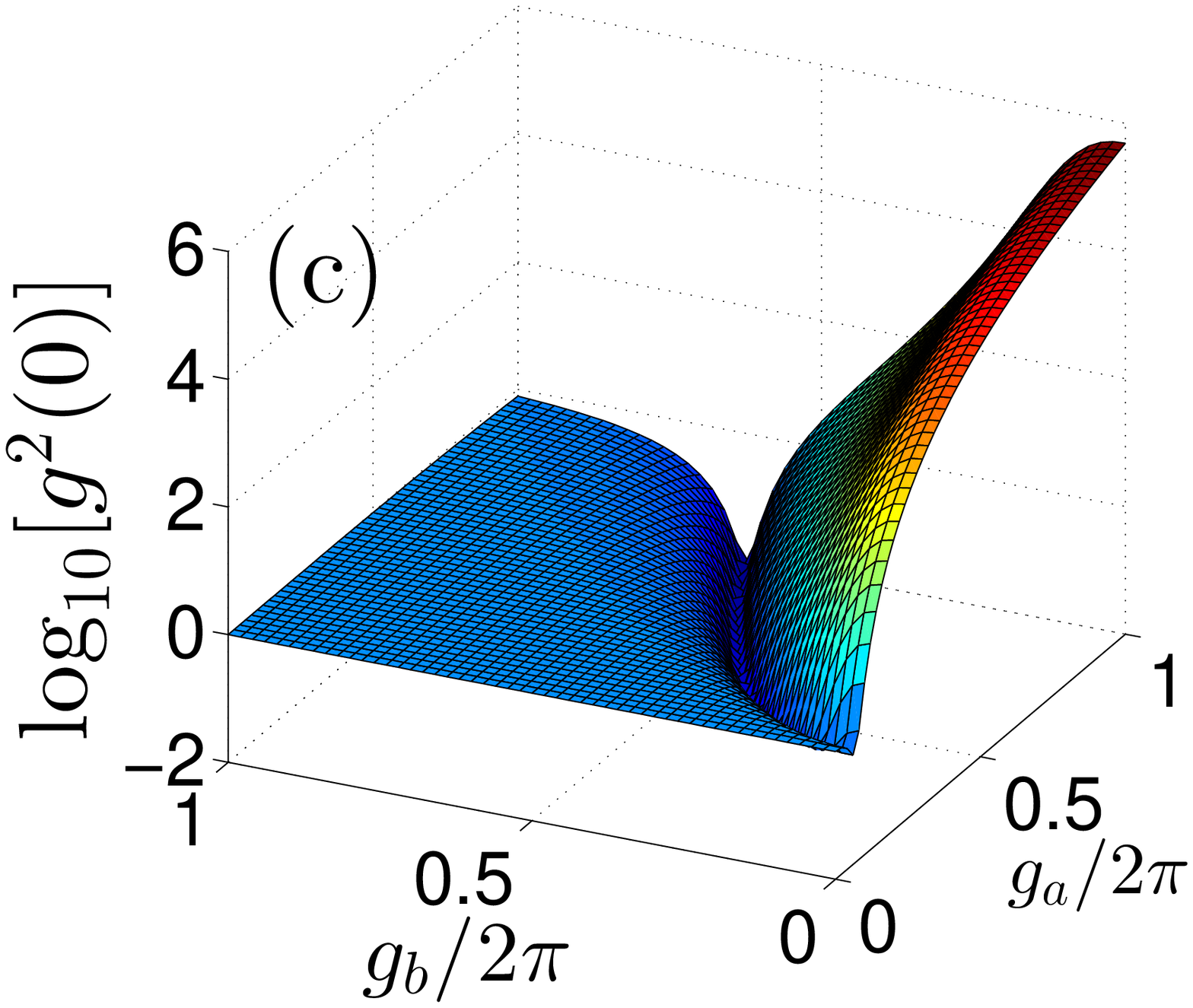}
\includegraphics[bb=10 173 577 664,  width=6cm, clip]{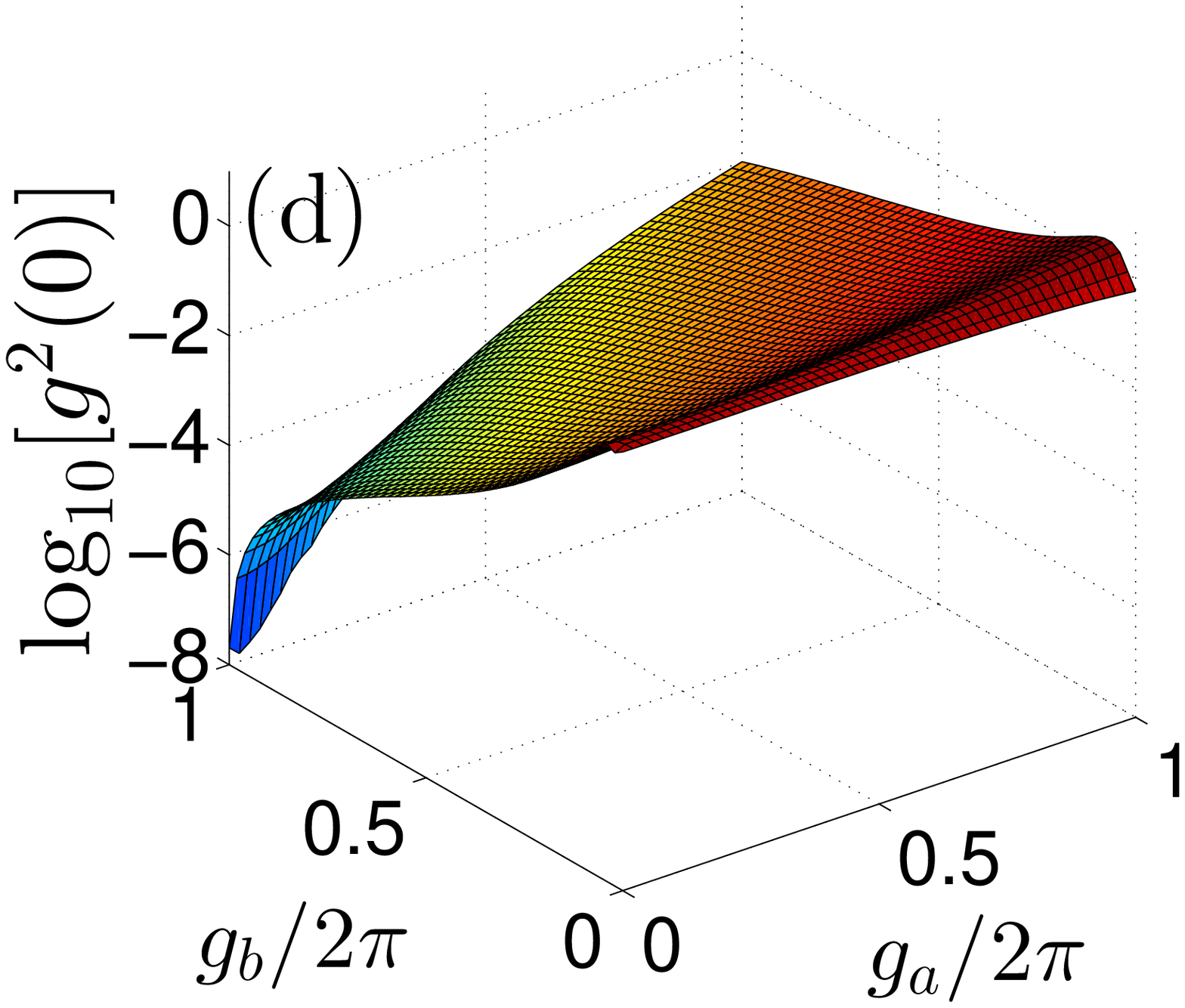}
\caption{(Color online) The second-order correlation functions
$g_{\mathrm{CCW}}^{\mathrm{(2)}}(0)$ of the CCW-propagating intercavity field are
plotted as functions of both cavity field detuning $\Delta/\kappa$ and
DQE-field coupling rate $g_{b}/\kappa$ (without mode coupling $J=0$
and with mode coupling $J=20\,\kappa$) for $g=0.5\,\kappa$, as shown
in (a) and (b), respectively. $g_{\mathrm{CCW}}^{\mathrm{(2)}}(0)$ is
plotted as functions of coupling strength $g_{\mathrm{a}}$ and $g_{\mathrm{b}}$
(without mode coupling $J=0$ and with mode coupling
$J=20\kappa$) for $\Delta=0$, as shown in (c) and (d), respectively. The
other system parameters are: $\kappa=40\gamma_{\mathrm{a}}$, $\varepsilon
=\gamma_{\mathrm{a}}$, and $\gamma_{\mathrm{a}}=1$ MHz.}%
\label{Figure15}%
\end{figure}

We then study the effect of the mismatch of the coupling strengths $g_{a}$ and
$g_{b}$ on the antibunching when the mode coupling is introduced. In
Fig.~\ref{Figure15}(a) and  Fig.~\ref{Figure15}(b), for given $g_{\mathrm{a}%
}=0.5\kappa$ and $\Delta=\Delta_{a}$, we plot $g_{\mathrm{CCW}}^{\mathrm{(2)}%
}(0)$ as functions of $\Delta$ and $g_{\mathrm{b}}$ with $J=0$ and $J=20\kappa$,
respectively. Fig.~\ref{Figure15}(a) shows that a small range around
$\Delta=0$ and $g_{\mathrm{b}}=0.4\kappa=0.8g_{\mathrm{a}}$ corresponds to
$g_{\mathrm{CCW}}^{\mathrm{(2)}}(0)<1$, which means photon antibunching.
However, in the other range, $g_{\mathrm{CCW}}^{\mathrm{(2)}}(0) \geq1$, which
corresponds to no photon antibunching. The minimum value appearing in
Fig.~\ref{Figure15}(a) for $J=0$ is very sensitive to $g_{\mathrm{b}}$ and
$\Delta$. We find that antibunching cannot be obtained even when the DQE-field
coupling strength $g_{b}$ becomes large. Thus a balance between
$g_{\mathrm{b}}$ and $g_{\mathrm{a}}$ is required to achieve photon
antibunching when there is no mode coupling.

When the mode coupling is introduced, e.g., $J=20\kappa$, Fig.~\ref{Figure15}%
(b) shows that the value of $g_{\mathrm{CCW}}^{\mathrm{(2)}}(0)$ first decreases
 and then increases very slowly when increasing $g_{b}$ for
$\Delta=0$. The minimum value of $g_{\mathrm{CCW}}^{\mathrm{(2)}}(0)$ is located
at $g_{\mathrm{b}}=\kappa$, which corresponds to the strongest antibunching. We
find that in the weak DQE-field coupling regime, i.e., $g_{\mathrm{b}%
}<\kappa$, the larger value of $g_{\mathrm{b}}$ corresponds to the smaller
value of $g_{\mathrm{CCW}}^{\mathrm{(2)}}(0)$ and then better antibunching.
However, in the strong-coupling regime, i.e., $g_{\mathrm{b}}>\kappa$, the
value of $g_{\mathrm{CCW}}^{\mathrm{(2)}}(0)$ increases with increasing
$g_{b}$, but the value of $g_{\mathrm{CCW}}^{\mathrm{(2)}}(0)$ is still very
small, i.e., $g_{\mathrm{CCW}}^{\mathrm{(2)}}(0)\leq0.0005$ in the given parameter regime.

To further find the effect of the mismatch between $g_{\mathrm{a}}$ and
$g_{\mathrm{b}}$ on the photon antibunching, $g_{\mathrm{CCW}}^{\mathrm{(2)}%
}(0)$ is plotted in Fig.~\ref{Figure15}(c) and in Fig.~\ref{Figure15}(d) as a
function of $g_{\mathrm{a}}$ and $g_{\mathrm{b}}$ for $\Delta=\Delta_{a}=0$,
with $J=0$ and $J \neq0$, e.g., $J=20\kappa$. For $J=0$, Fig.~\ref{Figure15}%
(c) clearly shows the antibunching can only occur in a narrow range which
needs a balance between $g_{\mathrm{a}}$ and $g_{\mathrm{b}}$ as previous
discussions. In most of the parameter range of $g_{a}$ and $g_{b}$, the value
of $g_{\mathrm{CCW}}^{\mathrm{(2)}}(0)$ satisfies the condition
$g_{\mathrm{CCW}}^{\mathrm{(2)}}(0)\geq1$, which means that there is no photon
antibunching. However, when mode coupling is introduced, e.g.,
$J=20\kappa$, the balance between $g_{\mathrm{a}}$ and $g_{\mathrm{b}}$ is not
required. Figure~\ref{Figure15}(d) also shows that the value of $g_{\mathrm{CCW}%
}^{\mathrm{(2)}}(0)$ increases when increasing both $g_{\mathrm{a}}$ and
$g_{b}$ for given other parameters. From Fig.~\ref{Figure15}(d), we also find a
very small value $10^{-8}$ of $g_{\mathrm{CCW} }^{\mathrm{(2)}}(0)$ can be
achieved at $g_{b}=\kappa$ when $g_{\mathrm{a}}=0$, i.e., the CCW mode is
decoupled from the DQE.

Therefore, we conclude that the strong robustness gainst mismatches discussed
above can be achieved when mode coupling is introduced. A balance, between
$g_{a}$ and $g_{b}$ for achieving better photon antibunching for the case
without mode coupling is not required when mode coupling is introduced.
In particular, we find that the mismatch between $g_{a}$ and $g_{b}$ can play
a positive role to produce photon antibunching when mode coupling is introduced.

\section{Conclusions}

In summary, we have studied photon antibunching in a system consisting of a
bimodal WGMC and a DQE. Two modes in the WGMC
are coupled via a scatterer. We mainly study the effect of the mode coupling on
the photon antibunching. it is known that photon antibunching cannot occur in
a system where a single-mode cavity field is weakly coupled to a DQE.
However, we find that a very strong antibunching can be achieved in the weak DQE-field
coupling regime, when mode coupling is introduced.

Comparing with the case when the DQE is coupled to two modes of the bimodal
cavity without mode coupling~\cite{23}, when mode coupling is introduced,
we find that: (i) a strong photon antibunching can be achieved in a larger parameter regime including both the weak and strong
DQE-field coupling regime; (ii) the minimum value of second-order
correlation function can be reduced up to several orders of magnitude even in the
weak DQE-field coupling regime; that is, a strong photon antibunching
can be achieved; (iii) the system is more robust to phase dephasing of the DQE; (iv)
the system is also more robust to the frequency mismatch between the cavity
modes and the DQE, as well as the mismatch of the coupling strengths between
two modes of the cavity and the DQE.

Our study shows that the system of two coupled cavity fields interacting with
a DQE could be a good quantum device for producing antibunched photons.
Our studies support a platform to achieve single-photon sources on
chip with robustness against fabrication imperfections and several
mismatches in the system.

\section{Acknowledgement}

Y.X.L. is supported by the National Basic Research Program of China 973 Program under Grant No.~2014CB921401, the National Natural Science Foundation of China under Grant Nos.~61328502, 61025022, the Tsinghua
University Initiative Scientific Research Program, and the Tsinghua National
Laboratory for Information Science and Technology (TNList) Cross-discipline Foundation.
FN is partly supported by the RIKEN iTHES Project, the MURI Center for Dynamic Magneto-Optics,
a Grant-in-Aid for Scientific Research (S), and the ImPACT Program of JST.

\appendix
\section{Analytical solutions of the second-order correlation
function with weak pump limit}

In the limit of a weak-driving field, the total excitation number of the
system is assumed no more than two. Using the ansatz given in Eqs. (22) and
(23) and combining with Eqs.~(24)-(26), the coefficients $C_{\mathrm{%
i,j,\pm }}$ in the steady state can be obtained via the Schr\"{o}dinger
equation by set $\partial \left\vert \varphi \right\rangle /\partial t=0$.
With the zero temperature approximation and neglecting the pure dephasing of
the DOE, we finally obtain the following set of linear equations

\begin{eqnarray}
0 &=&(\Delta -i\frac{\kappa }{2})C_{1,0,-}+JC_{0,1,-}+gC_{0,0,+}+\varepsilon
\text{,}  \label{AA1} \\
0 &=&(\Delta -i\frac{\kappa }{2})C_{0,1,-}+JC_{1,0,-}+gC_{0,0,+}\text{,}
\label{AA2} \\
0 &=&(\Delta -i\frac{\gamma _{a}}{2})C_{0,0,+}+gC_{1,0,-}+gC_{0,1,-}\text{,}
\label{AA3} \\
0 &=&(2\Delta -i\kappa )C_{2,0,-}+\sqrt{2}JC_{1,1,-}  \notag \\
&&+\sqrt{2}gC_{1,0,+}+\sqrt{2}\varepsilon C_{1,0,-}\text{,}  \label{AA4} \\
0 &=&(2\Delta -i\kappa )C_{0,2,-}+\sqrt{2}JC_{1,1,-}+\sqrt{2}gC_{0,1,+}\text{%
,}  \label{AA5} \\
0 &=&(2\Delta -i\kappa )C_{1,1,-}+\sqrt{2}JC_{0,2,-}+\sqrt{2}JC_{2,0,-}
\notag \\
&&+gC_{0,1,+}+gC_{1,0,+}+\varepsilon C_{0,1,-}\text{,}  \label{AA6} \\
0 &=&\left[ 2\Delta -i(\frac{\kappa +\gamma _{a}}{2})\right]
C_{1,0,+}+JC_{0,1,+}  \notag \\
&&+\sqrt{2}gC_{2,0,-}+gC_{1,1,-}+\varepsilon C_{0,0,+}\text{,}  \label{AA7}
\\
0 &=&\left[ 2\Delta -i(\frac{\kappa +\gamma _{a}}{2})\right]
C_{0,1,+}+JC_{1,0,+}  \notag \\
&&+gC_{1,1,-}+\sqrt{2}gC_{0,2,-}\text{.}  \label{AA8}
\end{eqnarray}%
Due to the weak pump limit $\varepsilon \rightarrow 0$, we can assume $%
C_{0,0,-}\rightarrow 1$, and one additional equation, namely $\varepsilon
C_{1,0,-}=0$, is irrelevant to the problem. The Eqs.~(\ref{AA1})-(\ref{AA8})
are now closed (i.e., eight equations for eight parameters). Thus, it is
possible to obtain the analytical solutions for all the coefficients $%
C_{i,j,\pm }$. However, the solution is cumbersome, so only $C_{1,0,-}$ and $%
C_{2,0,-}$ are given

\begin{eqnarray}
C_{1,0,-} &=&\frac{\varepsilon (\Delta _{\mathrm{p}}\Delta _{\mathrm{d}%
}-g^{2})}{(J-\Delta _{\mathrm{p}})C_{1}},  \label{AA9} \\
C_{2,0,-} &=&\frac{\sqrt{2}\varepsilon ^{2}L}{C_{3}},  \label{AA10}
\end{eqnarray}%
where%
\begin{eqnarray}
C_{1} &=&-2g^{2}+J\Delta _{\mathrm{d}}+\Delta _{\mathrm{p}}\Delta _{\mathrm{d%
}},  \label{AA11} \\
C_{2} &=&(J+\Delta _{\mathrm{p}})^{2}+\Delta _{\mathrm{d}}(J+\Delta _{%
\mathrm{p}})-2g^{2},  \label{AA12} \\
C_{3} &=&2(J-\Delta _{\mathrm{p}})^{2}C_{1}C_{2},  \label{AA13}
\end{eqnarray}%
\begin{eqnarray}
L &=&\Delta _{\mathrm{p}}^{3}\Delta _{\mathrm{d}}+\Delta _{\mathrm{p}%
}^{2}\Delta _{\mathrm{d}}^{2}+J\Delta _{\mathrm{p}}^{2}\Delta _{\mathrm{d}}\nonumber
\\
&&-2\Delta _{\mathrm{p}}\Delta _{\mathrm{d}}g^{2}-2J\Delta _{\mathrm{p}%
}g^{2}+g^{4},  \label{AA14} \\
\Delta _{\mathrm{p}} &=&\Delta -i\frac{\kappa }{2},  \label{AA15} \\
\Delta _{\mathrm{d}} &=&\Delta -i\frac{\gamma _{\mathrm{a}}}{2}.
\label{AA16}
\end{eqnarray}%
We substitute Eqs.~(\ref{AA9}) and (\ref{AA10}) into Eq.~(24), and then
can derive the analytical solutions of the $g_{\mathrm{CCW}}^{\mathrm{(2)}%
}(0)$ in the form of Eq.~(27).

\end{document}